\documentclass[a4paper,11pt]{article}
\usepackage{amsfonts}
\usepackage{amsmath}
\usepackage{amssymb}
\usepackage{mathtools}
\usepackage{colortbl}
\usepackage{cite}
\usepackage{hyperref}
\usepackage{textcomp}
\numberwithin{equation}{section}
\usepackage[left=2cm,right=2cm,top=3.5cm,bottom=3.5cm]{geometry}

\allowdisplaybreaks[1]
\hypersetup{
colorlinks=true,
linkcolor=ceruleanblue,
filecolor=ceruleanblue,      
urlcolor=ceruleanblue,
citecolor=ceruleanblue,
}
\definecolor{ceruleanblue}{rgb}{0.0, 0.2, 0.6}

\date{\today}

\begin{document}

\begin{flushright} {\footnotesize YITP-22-78, IPMU22-0039}  \end{flushright}

\begin{center}
\LARGE{\bf Generalized Regge-Wheeler Equation from \texorpdfstring{\\}{}
Effective Field Theory of Black Hole Perturbations\texorpdfstring{\\}{}
with a Timelike Scalar Profile}
\\[1cm] 

\large{Shinji Mukohyama$^{\,\rm a, \rm b}$, Kazufumi Takahashi$^{\,\rm a}$, and Vicharit Yingcharoenrat$^{\,\rm b}$}
\\[0.5cm]

\small{
\textit{$^{\rm a}$
Center for Gravitational Physics and Quantum Information, Yukawa Institute for Theoretical Physics, 
\\ Kyoto University, 606-8502, Kyoto, Japan}}
\vspace{.2cm}

\small{
\textit{$^{\rm b}$
Kavli Institute for the Physics and Mathematics of the Universe (WPI), The University of Tokyo Institutes for Advanced Study (UTIAS), The University of Tokyo, Kashiwa, Chiba 277-8583, Japan}}
\vspace{.2cm}
\end{center}

\vspace{0.3cm} 

\begin{abstract}\normalsize
Recently, the Effective Field Theory (EFT) of perturbations on an arbitrary background metric with a timelike scalar profile was formulated in the context of scalar-tensor theories. Here, we generalize the dictionary between the coefficients in the EFT action and those in covariant theories to accommodate shift- and reflection-symmetric quadratic higher-order scalar-tensor theories, including DHOST as well as U-DHOST. We then use the EFT action to study the dynamics of odd-parity perturbations on a static and spherically symmetric black hole background with a timelike scalar profile. Finally, we obtain the generalized Regge-Wheeler equation that can be used, e.g., to determine the spectrum of quasinormal modes and tidal Love numbers. 
\end{abstract}

\vspace{0.3cm} 

\vspace{2cm}

\newpage
{
	\hypersetup{linkcolor=black}
	\tableofcontents
}

\flushbottom

\vspace{1cm}


\section{Introduction}
Scalar-tensor gravity is a prototype of modified gravity theories and has a relatively long history, starting with Jordan in 1955~\cite{jordan1955schwerkraft} and Brans-Dicke in 1961~\cite{Brans:1961sx}. Further developments include discoveries of Horndeski's theory in 1974~\cite{Horndeski:1974wa} (and its rediscovery in the context of generalized Galileons in 2011~\cite{Deffayet:2011gz,Kobayashi:2011nu}), Degenerate Higher-Order Scalar-Tensor (DHOST) theories in 2015--2016~\cite{Langlois:2015cwa,Crisostomi:2016czh,BenAchour:2016fzp} (for comprehensive reviews, see \cite{Langlois:2018dxi,Kobayashi:2019hrl} and references therein), and U-DHOST theories in 2018~\cite{DeFelice:2018ewo} (see also \cite{DeFelice:2021hps,DeFelice:2022xvq}).\footnote{Also, one can use the so-called (invertible) disformal transformation~\cite{Bekenstein:1992pj,Bruneton:2007si,Bettoni:2013diz} or its higher-derivative extension~\cite{Takahashi:2021ttd} to generate new scalar-tensor theories from known ones~\cite{Zumalacarregui:2013pma,Domenech:2015tca,Takahashi:2017zgr,Takahashi:2021ttd,Takahashi:2022mew}.} Scalar-tensor gravity appears also as the low-energy limit of higher-dimensional theories such as string theory. Upon either compactifying extra dimension(s) or confining matter field(s) on a lower-dimensional brane, effective four-dimensional gravity can emerge at low energy under suitable conditions. Such effective four-dimensional gravity is often described by scalar-tensor theories with scalar field(s) stemming from the shape/size of extra dimensions or/and the position/bending of the brane in extra dimensions. 

In \cite{Mukohyama:2022enj}, the effective field theory (EFT) of scalar-tensor gravity with a single scalar field was formulated on an arbitrary background with a timelike scalar profile, by extending the EFT of ghost condensation~\cite{ArkaniHamed:2003uy,Arkani-Hamed:2003juy} and that of inflation/dark energy~\cite{Creminelli:2006xe,Cheung:2007st,Gubitosi:2012hu}. Once a background with a timelike scalar profile is specified, this EFT can universally describe perturbations in all scalar-tensor theories including (but not restricted to) the above mentioned known theories. This EFT indeed is motivated by many cosmological models such as inflation and dark energy, where the scalar field has a timelike gradient. Moreover, if one hopes to learn something about scalar-field dark energy or cosmological scalar-tensor gravity from astrophysical black holes, then one needs to consider black hole solutions with a timelike scalar profile.\footnote{There are also earlier works on EFTs of perturbations on black holes with a spacelike scalar profile~\cite{Franciolini:2018uyq,Hui:2021cpm}.} Of course, one could consider two independent EFTs: one valid on cosmological scales and the other on black hole scales. However, in this case, there is no known relation between parameters of one EFT and those of the other EFT (unless a suitable UV completion is found). This fact motivated two of the authors to formulate the EFT of perturbations~\cite{Mukohyama:2022enj} that can be applied to both black holes and cosmology. With this formulation, we hope to extract knowledge about scalar-field dark energy or cosmological scalar-tensor gravity from observations of astrophysical black holes. Furthermore, once a concrete theory is given, one can easily find the dictionary between the coefficients in the EFT action and the theory parameters, as explicitly shown in \cite{Mukohyama:2022enj} for the case of Horndeski's theory. Therefore, this EFT is expected to play as important a role in the context of black hole physics in scalar-tensor gravity as the parameterized post-Newtonian formalism~\cite{Will:2014xja} did in the context of solar system experiments of gravity. 

The main purposes of the present paper is to demonstrate a simple application of the general EFT developed in \cite{Mukohyama:2022enj} to a black hole background. In particular, we shall study odd-parity perturbations around a static and spherically symmetric black hole with a timelike scalar profile and obtain the generalized Regge-Wheeler equation. There have been extensive studies on perturbations of black holes with a timelike scalar profile in Horndeski/DHOST theories~\cite{Ogawa:2015pea,Takahashi:2015pad,Takahashi:2016dnv,Tretyakova:2017lyg,Babichev:2017lmw,Babichev:2018uiw,Minamitsuji:2018vuw,Takahashi:2019oxz,deRham:2019gha,Charmousis:2019fre,Khoury:2020aya,Tomikawa:2021pca,Langlois:2021aji,Langlois:2021xzq,Takahashi:2021bml,Nakashi:2022wdg,Langlois:2022ulw}, and our EFT framework encompasses and extends\footnote{The EFT includes the so-called scordatura terms~\cite{Motohashi:2019ymr} that are absent in DHOST theories but that are present in ghost condensation~\cite{Arkani-Hamed:2003pdi,Mukohyama:2005rw}.} these works in principle. Another purpose of the paper is to extend the dictionary between the EFT coefficients and parameters in concrete covariant theories to general quadratic higher-order scalar-tensor theories including DHOST and U-DHOST theories, under the assumption of shift and reflection symmetries. 

The rest of the present paper is organized as follows. In Section~\ref{sec:overview}, we give a brief review on the construction of the EFT of perturbations on an arbitrary background metric with a timelike scalar profile, following \cite{Mukohyama:2022enj}. In Section~\ref{sec:dictionary}, we report a dictionary between the EFT and the shift- and $Z_2$-symmetric higher-order scalar-tensor theories. In Section~\ref{sec:background}, we restrict ourselves to the class of static and spherically symmetric background geometries, and find the background equations of motion derived from the EFT action. In Section~\ref{sec:odd-parity}, we use the EFT action to analyze the dynamics of odd-parity perturbations. Finally, we conclude the paper in Section~\ref{sec:conclusions}.


\section{Overview of the EFT}\label{sec:overview}
In this Section, we briefly review the construction of the EFT of perturbations on an arbitrary background metric with a timelike scalar profile, following \cite{Mukohyama:2022enj}. The basic idea is that one chooses the background of the scalar field, $\bar{\Phi}$, to be time-dependent which spontaneously breaks the time diffeomorphism. On top of that, it is important to make sure that such a time-dependent scalar field defines a proper $3+1$ foliation of a spacetime manifold i.e. the background of $\Phi$ determines the time direction. Therefore, the unbroken symmetries on a constant-$\Phi$ hypersurface are the 3d diffeomorphism. We denote the time coordinate by $\tau$ and then define a unit vector normal to the hypersurface, 
\begin{align}\label{eq:normal}
	n_\mu \equiv - \frac{\partial_\mu \Phi}{\sqrt{-X}} \rightarrow -\frac{\delta^\tau_\mu}{\sqrt{-g^{\tau\tau}}} \;,
\end{align}
where $g^{\tau\tau}$ denotes the ($\tau\tau$)-component of the inverse metric, $X = g^{\mu\nu}\partial_\mu \Phi \partial_\nu \Phi<0$, and the vector~$n^\mu$ satisfies $n_\mu n^\mu = -1$ and $n^{\mu}\partial_{\mu}\Phi=\sqrt{-X}>0$. The arrow refers to the expression in the unitary gauge where $\Phi=\bar{\Phi}(\tau)$. It is convenient to work with the Arnowitt-Deser-Misner (ADM) decomposition in which the metric is written in the form,
\begin{align}\label{eq:metric_ADM}
	ds^2 = - N^2 d\tau^2 + h_{ij}(dx^i + N^i d\tau)(dx^j + N^j d\tau) \;,
\end{align}
where $N$ is the lapse function, $N^i$ is the shift vector, and $h_{\mu\nu} = g_{\mu\nu} + n_\mu n_\nu$ is the induced metric. On a constant-$\Phi$ hypersurface, the extrinsic curvature is defined by
\begin{align}\label{eq:extrinsic}
	K_{\mu\nu}\equiv h_\mu^\alpha\nabla_\alpha n_\nu \;,
\end{align}
where $\nabla_\mu$ is the 4d covariant derivative. In terms of the ADM variables introduced in (\ref{eq:metric_ADM}), the extrinsic curvature and its trace are given by 
\begin{align}
	K_{ij} = \frac{1}{2N}(\dot{h}_{ij} - D_i N_j - D_j N_i) \;, \quad K = h^{ij}K_{ij} \;,
\end{align}
where $D_i$ is the covariant derivative associated with the induced metric~$h_{ij}$ and a dot denotes the derivative with respect to $\tau$. Furthermore, one can straightforwardly construct the 3d Ricci tensor~${}^{(3)}\!R_{ij}$ and its trace~${}^{(3)}\!R$ using the induced metric and the 3d Christoffel symbol. 

Following the same procedure as it has been done for the EFT of dark energy/inflation~\cite{Cheung:2007st,Gubitosi:2012hu}, one can write down an EFT action in the unitary gauge~$\Phi=\bar{\Phi}(\tau)$ which contains terms invariant under the 3d diffeomorphism,
\begin{align}\label{eq:unitary_action}
	S = \int d^4x \sqrt{-g}\,F(\tilde{R}_{\mu\nu\alpha\beta},g^{\tau\tau},K_{\mu\nu},\nabla_\nu,\tau) \;.
\end{align}
Here, $\tilde{R}_{\mu\nu\alpha\beta}$ is the 4d Riemann tensor and $F$ is a scalar function under the 3d diffeomorphism made out of geometrical quantities. We see that this action can contain not only terms that are invariant under the 4d diffeomorphism such as the 4d Ricci scalar~$\tilde{R}$ but also terms built out of, for instance, $N$, $K_{\mu\nu}$, and ${}^{(3)}\!R$ that are manifestly invariant only under the 3d diffeomorphism. Furthermore, the fact that the $\tau$-diffeomorphism is spontaneously broken by $\bar{\Phi}(\tau)$ allows the action to contain explicitly $\tau$-dependent functions. 

Although the action~(\ref{eq:unitary_action}) in general can be applied to generic background geometries without assuming any symmetries, it is not yet the EFT of perturbations we wanted to achieve. As we will see below, we will expand the action~(\ref{eq:unitary_action}) in terms of perturbations and we will impose a set of consistency relations on the expansion coefficients to make sure that the 3d diffeomorphism invariance is preserved as a whole. 

Let us now define the perturbations of $g^{\tau\tau}$, the extrinsic curvature, and the 3d Ricci tensor as follows:
\begin{align}\label{eq:building_blocks}
	\delta g^{\tau\tau} \equiv g^{\tau\tau} - \bar{g}^{\tau\tau}(\tau,\vec{x}) \;, \quad \delta K^\mu_\nu \equiv K^{\mu}_\nu - \bar{K}^\mu_\nu(\tau,\vec{x}) \;, \quad  \delta {}^{(3)}\!R^\mu_\nu \equiv {}^{(3)}\!R^\mu_\nu - {}^{(3)}\!\bar{R}^\mu_\nu(\tau,\vec{x}) \;,
\end{align}
where the background values are denoted with a bar and $\vec{x}$ refers to spatial coordinates. Notice that the background quantities as defined above can generically depend on the spatial coordinates.

As pointed out in \cite{Mukohyama:2022enj}, the naive EFT action which is written in terms of the building blocks such as $\delta g^{\tau\tau}$ and $\delta K$ is not invariant under the 3d diffeomorphism due to the presence of the background quantities, e.g., $\bar{g}^{\tau\tau}$, $\bar{K}$, and ${}^{(3)}\!\bar{R}$ that are in general functions of spatial coordinates. In order to achieve a consistent EFT action, one then Taylor expands the unitary gauge action~(\ref{eq:unitary_action}), that is manifestly invariant under the 3d diffeomorphism, in terms of perturbations defined in (\ref{eq:building_blocks}), 
\begin{align}
	S = \int d^4x \sqrt{-g}~\bigg[&\bar{F} + \bar{F}_{g^{\tau\tau}} \delta g^{\tau\tau} + \bar{F}_{K}\delta K + \bar{F}_{\sigma^\mu_\nu}\delta \sigma^\mu_\nu + \bar{F}_{{}^{(3)}\!R}\delta {}^{(3)}\!R +  \bar{F}_{r^\mu_\nu}\delta r^\mu_\nu + \bar{F}_{n^\nu\partial_\nu g^{\tau\tau}} \bar{n}^\mu \partial_\mu \delta g^{\tau\tau} \nonumber \\ 
	&+ \frac{1}{2}\bar{F}_{g^{\tau\tau}g^{\tau\tau}}(\delta g^{\tau\tau})^2 + \bar{F}_{g^{\tau\tau}K} \delta g^{\tau\tau}\delta K + \bar{F}_{g^{\tau\tau} \sigma^\mu_\nu}\delta \sigma^\mu_\nu \delta g^{\tau\tau}  + \bar{F}_{g^{\tau\tau} r^\mu_\nu}\delta r^\mu_\nu \delta g^{\tau\tau}  \nonumber \\
	& +  \bar{F}_{g^{\tau\tau}{}^{(3)}\!R}\delta g^{\tau\tau} \delta {}^{(3)}\!R  + \frac{1}{2}\bar{F}_{KK} \delta K^2 + \bar{F}_{K \sigma^\mu_\nu}\delta K \delta \sigma^\mu_\nu + \bar{F}_{K{}^{(3)}\!R}\delta K \delta {}^{(3)}\!R + \bar{F}_{K r^\mu_\nu}\delta K \delta r^\mu_\nu    \nonumber \\ 
	&  + \frac{1}{2} \bar{F}_{{}^{(3)}\!R{}^{(3)}\!R}\delta {}^{(3)}\!R^2   + \bar{F}_{{}^{(3)}\!R \sigma^\mu_\nu} \delta {}^{(3)}\!R \delta \sigma^\mu_\nu  +   \bar{F}_{{}^{(3)}\!R r^\mu_\nu}\delta {}^{(3)}\!R \delta r^\mu_\nu+\frac{1}{2}\bar{F}_{r^2 }\delta r^\mu_\nu \delta r^\nu_\mu   \nonumber \\
	& + \frac{1}{2}\bar{F}_{\sigma^2} \delta \sigma^\mu_\nu \delta \sigma^\nu_\mu + \bar{F}_{\sigma r }\delta \sigma^\mu_\nu \delta r^\nu_\mu + \frac{1}{2} \bar{F}_{(n^\nu \partial_\nu g^{\tau\tau})^2}(\bar{n}^\mu \partial_\mu \delta g^{\tau\tau})^2 + \bar{F}_{K n^\mu \partial_\mu g^{\tau\tau}} \delta K (\bar{n}^\mu \partial_\mu \delta g^{\tau\tau}) \nonumber \\
	& + \frac{1}{2} \bar{F}_{h^{\mu\nu}\partial_\mu g^{\tau\tau} \partial_\nu g^{\tau\tau}} \bar{h}^{\mu\nu} \partial_\mu \delta g^{\tau\tau} \partial_\nu \delta g^{\tau\tau} + \cdots \bigg] \label{eq:action_taylor}\;,
\end{align}
where $\sigma_{\mu\nu} \equiv K_{\mu\nu} - K h_{\mu\nu}/3$, $r_{\mu\nu} \equiv {}^{(3)}\!R_{\mu\nu} - {}^{(3)}\!R h_{\mu\nu}/3$, and we have expanded the action up to second order in perturbations. We use $\bar{F}_{X} \equiv (\partial F/\partial X)_{\rm BG}$ etc.\ to denote the Taylor coefficients evaluated on the background. Clearly, we see that each term of the action above breaks 3d diffeomorphism due to the spatial dependence of the background quantities. However, since the full action must be invariant under the 3d diffeomorphism, there must be a set of consistency relations that guarantees the 3d diffeomorphism invariance of the action as a whole. 

As shown in \cite{Mukohyama:2022enj}, a set of consistency relations can be simply obtained by applying the chain rule to each term of the expansion. For example, the chain rule associated with spatial derivatives of $\bar{F}(\tau,\vec{x})$ is
\begin{align}
	\frac{\partial}{\partial x^i}\bar{F}(\tau,\vec{x}) &= \frac{d}{dx^i} F(g^{\tau\tau},K,{}^{(3)}\!R,\tau) \bigg|_{\rm BG} \nonumber \\
	&= \bar{F}_{g^{\tau\tau}}\frac{\partial \bar{g}^{\tau\tau}}{\partial x^i} + \bar{F}_{K}\frac{\partial \bar{K}}{\partial x^i} + \bar{F}_{\sigma^\mu_\nu} \frac{\partial \bar{\sigma}^\mu_\nu}{\partial x^i} + \bar{F}_{{}^{(3)}\!R} \frac{\partial {}^{(3)}\!\bar{R}}{\partial x^i} + \bar{F}_{r^\mu_\nu} \frac{\partial \bar{r}^\mu_\nu}{\partial x^i} +  \bar{F}_{n^\nu\partial_\nu g^{\tau\tau}} \frac{\partial(\bar{n}^\mu\partial_\mu \bar{g}^{\tau\tau})}{\partial x^i} \label{eq:con2}\;,
\end{align}
where we have omitted terms of higher order in derivatives. In principle, there are infinitely many consistency relations which can be obtained by applying the chain rule to other terms in (\ref{eq:action_taylor}) such as $\bar{F}_{g^{\tau\tau}}$ and $\bar{F}_K$ (see \cite{Mukohyama:2022enj} for more detail). Notice that the chain rule associated to $\tau$-derivative does not lead to non-trivial relations among the EFT coefficients since the chain rule in this case involves $F_{\tau}$, which does not show up as an EFT coefficient. In addition, as expected, such consistency relations are automatically satisfied in covariant theories, for example, Horndeski and (U-)DHOST theories (see also Section~\ref{sec:dictionary}). 

We now write down the EFT action in the unitary gauge, up to second order in perturbations, 
\begin{align}
	S = \int d^4x &\sqrt{-g} \bigg[\frac{M_\star^2}{2}f(y)R - \Lambda(y) - c(y)g^{\tau\tau} - \beta(y) K  - \alpha^\mu_{\nu}(y)\sigma^{\nu}_\mu - \gamma^\mu_{\nu}(y)r^{\nu}_\mu - \zeta(y) \bar{n}^\mu\partial_\mu g^{\tau\tau}  \nonumber \\ 
	&+ \frac{1}{2} m_2^4(y) (\delta g^{\tau\tau})^2  + \frac{1}{2} M_1^3(y) \delta g^{\tau\tau} \delta K + \frac{1}{2} M_2^2(y) \delta K^2 + \frac{1}{2} M_3^2(y) \delta K^\mu_\nu \delta K^\nu_\mu  \nonumber \\
	& + \frac{1}{2}M_4(y) \delta K \delta {}^{(3)}\!R + \frac{1}{2}M_5(y) \delta K^\mu_\nu \delta {}^{(3)}\!R^\nu_\mu + \frac{1}{2}\mu_1^2(y) \delta g^{\tau\tau} \delta {}^{(3)}\!R + \frac{1}{2}\mu_2(y)\delta {}^{(3)}\!R^2 \nonumber \\
	&+ \frac{1}{2} \mu_3(y) \delta {}^{(3)}\!R^\mu_\nu \delta {}^{(3)}\!R^\nu_\mu  + \frac{1}{2} \lambda_1(y)^\nu_\mu \delta g^{\tau\tau} \delta K^\mu_\nu + \frac{1}{2} \lambda_2(y)^\nu_\mu \delta g^{\tau\tau} \delta {}^{(3)}\!R^\mu_\nu +  \frac{1}{2} \lambda_3(y)^\nu_\mu \delta K \delta K^\mu_\nu  \nonumber \\ 
	&+  \frac{1}{2} \lambda_4(y)^\nu_\mu \delta K \delta {}^{(3)}\!R^\mu_\nu + \frac{1}{2} \lambda_5(y)^\nu_\mu \delta {}^{(3)}\!R \delta K^\mu_\nu + \frac{1}{2} \lambda_6(y)^\nu_\mu \delta {}^{(3)}\!R \delta {}^{(3)}\!R^\mu_\nu + \frac{1}{2}{\mathcal M}_1^2(y)(\bar{n}^\mu\partial_\mu\delta g^{\tau\tau})^2  \nonumber \\
	&+ \frac{1}{2}{\mathcal M}_2^2(y)\delta K(\bar{n}^\mu\partial_\mu\delta g^{\tau\tau})+\frac{1}{2}{\mathcal M}_3^2(y)\bar{h}^{\mu\nu}\partial_\mu\delta g^{\tau\tau}\partial_\nu\delta g^{\tau\tau} + \cdots \bigg] \;, \label{eq:EFT}
\end{align}
where $y = \{\tau, \vec{x}\}$ and the ellipsis refers to higher-order operators.
The terms in the first line are tadpole terms and (a part of) the functions~$f(y)$, $\Lambda(y)$, $c(y)$, $\beta(y)$, $\alpha^\mu_\nu(y)$, $\gamma^\mu_\nu(y)$, and $\zeta(y)$ will be fixed by the background equations of motion (see Section~\ref{sec:background} for the case of static and spherically symmetric background metric).
Here and in what follows, for simplicity, we work with the 4d Ricci scalar with the boundary term subtracted, 
\begin{align} \label{eqn:def-R}
	R &\equiv {}^{(3)}\!R + K_{\mu\nu}K^{\mu\nu} - K^2 = \tilde{R} - 2\nabla_\mu( K n^\mu - n^\nu\nabla_\nu n^\mu)\;.
\end{align}
Notice that, compared with Eq.~(2.21) of \cite{Mukohyama:2022enj}, there are four additional terms that contain the derivative of $\delta g^{\tau\tau}$. As we will see in Section~\ref{sec:dictionary}, these extra terms in fact correspond to the functions characterizing quadratic higher-order scalar-tensor theories. Here, we only report the relations between the extra parameters in (\ref{eq:EFT}) and the Taylor coefficients in (\ref{eq:action_taylor}): 
\begin{align}\label{eq:EFT_para_extra}
	\zeta(y) = -\bar{F}_{n^\mu \partial_\mu g^{\tau\tau}} \;, \quad \mathcal{M}_1^2 = \bar{F}_{(n^\mu \partial_\mu g^{\tau\tau})^2} \;, \quad \mathcal{M}_2^2 = 2 \bar{F}_{K n^\mu \partial_\mu g^{\tau\tau}}\;, \quad \mathcal{M}_3^2 = 
	\bar{F}_{h^{\mu\nu}\partial_\mu g^{\tau\tau} \partial_\nu g^{\tau\tau}}\;.
\end{align}
The other relations between the EFT parameters and the Taylor coefficients are the same as Eq.~(2.23) of \cite{Mukohyama:2022enj}, except there is an extra contribution~$\bar{F}_{n^\nu \partial_\nu g^{\tau\tau}}\bar{n}^\mu \partial_\mu \bar{g}^{\tau\tau}$ to the expression of $\Lambda(y)$. Moreover, the consistency relations in terms of the EFT parameters can be easily obtained using (\ref{eq:EFT_para_extra}) above and Eq.~(2.23) of \cite{Mukohyama:2022enj}. For instance, from (\ref{eq:con2}), we have 
\begin{align}
	\partial_i\Lambda + \bar{g}^{\tau\tau}\partial_i c - \frac{1}{2}M^2_\star {}^{(3)}\!\bar{R} \partial_i f + \frac{1}{3}\bar{K}(M^2_\star \bar{K}\partial_i f + 3\partial_i\beta) - \frac{1}{2} \bar{\sigma}^\mu_\nu &(M^2_\star \bar{\sigma}^\nu_\mu \partial_i f  - 2\partial_i\alpha^\nu_\mu) \nonumber \\
	&+ \bar{r}^\mu_\nu\partial_i\gamma^{\nu}_\mu + \bar{n}^\mu \partial_\mu \bar{g}^{\tau\tau} \partial_i \zeta \simeq 0  \label{eq:con_EFT1} \;,
\end{align}
where the symbol~$\simeq$ means that we have omitted subleading terms suppressed under an appropriate scaling (see the discussion below).
Note that the last term in the equation above is due to the operator~$\zeta(y) \bar{n}^\mu\partial_\mu g^{\tau\tau}$ in the EFT action~(\ref{eq:EFT}), which is newly introduced in the present paper. Additionally, the other consistency relations can be straightforwardly obtained following the same method as explained above. 

Our EFT can be easily extended to incorporate matter fields whose background profile can depend on spatial coordinates and which can source inhomogeneities of the background spacetime (e.g., the case of neutron stars).
In this case, we just start from the EFT action~\eqref{eq:EFT} supplemented with a 4d covariant matter action minimally coupled to the metric $g_{\mu\nu}$.\footnote{It is reasonable to assume that all the matter fields (at least in the visible sector) are coupled to the same (Jordan-frame) metric so as to respect the weak equivalence principle. Also, one could perform a conformal/disformal transformation to move to the Einstein frame ($f=1$) at the price of non-minimal coupling in the matter sector. In the case of black hole where we do not need to specify how matter fields are coupled to gravity, moving to the Einstein frame greatly simplifies the analysis (see Section~\ref{sec:background}).}
Then, the tadpole cancellation conditions yield the background equations of motion in the presence of matter fields.
The form of the consistency relations remains the same as far as they are formulated before one imposes the tadpole cancellation conditions. 
Having said this, our main focus in this paper lies on black hole (i.e., vacuum) backgrounds, and hence we disregard matter fields in our EFT action.

Before ending this Section, let us comment on the relevant scales of the four extra terms in (\ref{eq:EFT}). As usual, we define an energy scale~$E$ that captures relevant scales of the background geometry: 
\begin{align}\label{eq:energy_E}
	E \equiv {\rm max}\{|{}^{(3)}\!\bar{R}|^{1/2} , |{}^{(3)}\!\bar{R}^{\mu}_{\nu}{}^{(3)}\!\bar{R}^{\nu}_{\mu}|^{1/4} , |\bar{K}| , |\bar{K}^{\mu}_{\nu}\bar{K}^{\nu}_{\mu}|^{1/2} , |{}^{(3)}\!\bar{R}^{\mu}_{\nu}\bar{K}^{\nu}_{\mu}|^{1/3}\} \;.
\end{align}
Also, we introduce the energy scales~$\mu$ and $\Lambda_\star$ to be a Lorentz breaking scale [energy scale for $\bar{\Phi}(\tau)$] and the cutoff of the EFT, respectively. Note that it is reasonable to assume that $\mu \gg \Lambda_\star$ and $E < \Lambda_\star$. Therefore, in terms of the relevant energy scales, we can define the scales associated with the four additional EFT coefficients as
\begin{align}
	\zeta \sim \mathcal{O}(M_\star^2 E^2 \Lambda_\star^{-1}) \;, \quad \mathcal{M}_1^2 \sim \mathcal{M}_3^2 \sim \mathcal{O}(M_\star^2 E^2 \Lambda_\star^{-2}) \;, \quad \mathcal{M}_2^2 \sim \mathcal{O}(M_\star^2 E \Lambda_\star^{-1})\;.
\end{align}
In the next Section, we will give a dictionary between the EFT coefficients defined in (\ref{eq:EFT}) and the functions in the covariant action of (shift- and reflection-symmetric) quadratic higher-order scalar-tensor theories.

\section{Dictionary}\label{sec:dictionary}

In this section, we discuss how the shift- and reflection-symmetric subclass of quadratic higher-order scalar-tensor theories described by the action
\begin{align}
	S=\int d^4x\sqrt{-g}\bigg[
	&P(X)+F(X)\tilde{R}+A_1(X)\nabla^\mu\nabla^\nu\Phi\nabla_\mu\nabla_\nu\Phi+A_2(X)(\Box\Phi)^2 \nonumber \\
	&+\frac{1}{2}A_3(X)\nabla^\mu\Phi\nabla_\mu X\Box\Phi+\frac{1}{4}A_4(X)\nabla^\mu X\nabla_\mu X+\frac{1}{4}A_5(X)(\nabla^\mu\Phi\nabla_\mu X)^2\bigg]\;, \label{HOST}
\end{align}
where the second derivative of $\Phi$ is contained up to the quadratic order, is embedded in our EFT.
(See \cite{Motohashi:2020wxj} for a similar discussion in the case of EFT of inflation.)
Here, $A_i$'s are arbitrary functions of the kinetic term of the scalar field~$X=\nabla^\mu\Phi\nabla_\mu\Phi$.
We do not impose any particular relation between the coefficient functions in \eqref{HOST} unless otherwise stated.

We use Eqs.~(\ref{eq:normal})--(\ref{eq:extrinsic}) and the acceleration vector given by
\begin{align}\label{eq:acceleration}
	a_\mu\equiv n^\alpha\nabla_\alpha n_\mu
	=-\frac{1}{2X}h_\mu^\alpha\partial_\alpha X \;,
\end{align}
which enable us to express the second covariant derivative of the scalar field as
\begin{equation}
	\nabla_\mu\nabla_\nu\Phi
	=-\sqrt{-X}\left(K_{\mu\nu}-2n_{(\mu}a_{\nu)}\right)-\frac{n^\alpha\partial_\alpha X}{2\sqrt{-X}}n_\mu n_\nu\;.
\end{equation}
Hence, the action~\eqref{HOST} takes the form
\begin{align}
	S=\int d^4x\sqrt{-g}\left[
	P+FR+f_1K^\mu_\nu K^\nu_\mu +f_2K^2+f_3(n^\mu\partial_\mu X)^2+f_4Kn^\mu\partial_\mu X+f_5a^\mu a_\mu
	\right]\;,
	\label{HOST2}
\end{align}
where $K\equiv g^{\mu\nu}K_{\mu\nu}$ and
\begin{equation}
	\begin{split}
		&f_1=-XA_1\;, \qquad
		f_2=-XA_2\;, \qquad
		f_3=-\frac{1}{4X}(A_1+A_2+XA_3+XA_4+X^2A_5)\;, \\
		&f_4=-\frac{1}{2}(2A_2+XA_3+4F_X)\;, \qquad
		f_5=X(2A_1+XA_4-4F_X)\;.
	\end{split} \label{eq:fs}
\end{equation}
Here, a subscript~$X$ denotes the derivative with respect to $X$. 

Assuming $\bar{\Phi}(\tau) = \mu^2\tau$ and $\bar{g}^{\tau\tau}=-1$ for simplicity (so that $\bar{X}=const.$) and expanding the action~\eqref{HOST2} up to second order in perturbations, we have
\begin{align}\label{eq:EFT_dictionary}
	S = \int d^4x \sqrt{-g} \bigg[&\frac{M_\star^2}{2}f(y)R - \Lambda(y) - c(y)g^{\tau\tau} - \tilde{\beta}(y) K - \alpha(y)\bar{K}^\mu_\nu K^{\nu}_\mu -\zeta(y) n^\mu\partial_\mu g^{\tau\tau} \nonumber \\ 
	& + \frac{1}{2} m_2^4(y) (\delta g^{\tau\tau})^2 + \frac{1}{2} \tilde{M}_1^3(y) \delta g^{\tau\tau} \delta K + \frac{1}{2} M_2^2(y) \delta K^2 + \frac{1}{2} M_3^2(y) \delta K^\mu_\nu \delta K^\nu_\mu \nonumber \\
	& + \frac{1}{2}\mu_1^2(y) \delta g^{\tau\tau} \delta {}^{(3)}\!R + \frac{1}{2} \lambda_1(y)^\mu_\nu \delta g^{\tau\tau} \delta K^\nu_\mu + \frac{1}{2}{\mathcal M}_1^2(y)(\bar{n}^\mu\partial_\mu\delta g^{\tau\tau})^2 \nonumber \\
	& +\frac{1}{2}{\mathcal M}_2^2(y)\delta K(\bar{n}^\mu\partial_\mu\delta g^{\tau\tau})+\frac{1}{2}{\mathcal M}_3^2(y)\bar{h}^{\mu\nu}\partial_\mu\delta g^{\tau\tau}\partial_\nu\delta g^{\tau\tau}+\cdots \bigg] \;,
\end{align}
with
\begin{equation}
	\begin{split}
		M_\star^2f&=2\bar{F}\;, \\
		\Lambda&=-\bar{P}+\bar{X}\bar{P}_X+\bar{X}\bar{F}_X{}^{(3)}\!\bar{R}+(\bar{X}\bar{F}_X+\bar{f}_1+\bar{X}\bar{f}_{1X})\bar{K}^\mu_\nu\bar{K}^\nu_\mu-(\bar{X}\bar{F}_X-\bar{f}_2-\bar{X}\bar{f}_{2X})\bar{K}^2\;, \\
		c&=\bar{X}\bar{P}_{X}+\bar{X}\bar{F}_X{}^{(3)}\!\bar{R}+\bar{X}(\bar{F}_X+\bar{f}_{1X})\bar{K}^\mu_\nu\bar{K}^\nu_\mu-\bar{X}(\bar{F}_X-\bar{f}_{2X})\bar{K}^2\;, \\
		\tilde{\beta}&=-2\bar{f}_2\bar{K}\;, \qquad
		\alpha=-2\bar{f}_1\;, \qquad
		\zeta=\bar{X}\bar{f}_4\bar{K}\;, \\
		m_2^4&=\bar{X}^2\bar{P}_{XX}+\bar{X}^2\bar{F}_{XX}{}^{(3)}\!\bar{R}+\bar{X}^2(\bar{F}_{XX}+\bar{f}_{1XX})\bar{K}^\mu_\nu\bar{K}^\nu_\mu-\bar{X}^2(\bar{F}_{XX}-\bar{f}_{2XX})\bar{K}^2 \\
		&\quad -\bar{X}^2\bar{f}_{4X}\bar{\nabla}_\mu(\bar{n}^\mu\bar{K})\;, \\
		\tilde{M}_1^3&=4\bar{X}(\bar{F}_X-\bar{f}_{2X})\bar{K}\;, \qquad
		M_2^2=2\bar{f}_2\;, \qquad
		M_3^2=2\bar{f}_1\;, \\
		\mu_1^2&=-2\bar{X}\bar{F}_X\;, \qquad
		\lambda_1{}^\mu_\nu=-4\bar{X}(\bar{F}_X+\bar{f}_{1X})\bar{K}^\mu_\nu\;, \\
		{\mathcal M}_1^2&=2\bar{X}^2\bar{f}_3\;, \qquad
		{\mathcal M}_2^2=-2\bar{X}\bar{f}_4\;, \qquad
		{\mathcal M}_3^2=\frac{1}{2}\bar{f}_5\;.
	\end{split} \label{EFT_coeff_HOST}
\end{equation}
Note that the action~\eqref{eq:EFT_dictionary} is a special case of the original EFT action~\eqref{eq:EFT}.
Here, for later convenience, we have employed a slightly different parameterization compared to the one in \eqref{eq:EFT}: 
The tadpole term~$\alpha^\mu_\nu(y)\sigma^\nu_\mu$ in \eqref{eq:EFT} is now replaced by $\alpha(y)\bar{K}^\mu_\nu K^\nu_\mu$ with a new EFT coefficient~$\alpha(y)$, and we express $R$ in terms of $K_{\mu\nu}$ instead of $\sigma_{\mu\nu}$.
Hence, we put a tilde on the coefficients which are modified due to this change (i.e., $\tilde{\beta} = \beta - \alpha \bar{K}/3$ and $\tilde{M}_1^3 = 3M_1^3/2$). 
Additionally, the consistency relations mentioned in Section~\ref{sec:overview} are automatically satisfied for the above choice of EFT coefficients.

It should be noted that so far we have not assumed any degeneracy condition on the higher-derivative terms in \eqref{HOST}, but the above dictionary of course applies to theories with degenerate higher-derivative terms, i.e., Horndeski theories~\cite{Horndeski:1974wa,Deffayet:2011gz,Kobayashi:2011nu}, DHOST theories~\cite{Langlois:2015cwa,Crisostomi:2016czh}, and U-DHOST theories~\cite{DeFelice:2018ewo,DeFelice:2021hps,DeFelice:2022xvq}.
In such specific theories, the degeneracy conditions impose some constraints on the EFT coefficients.
In particular, the shift- and reflection-symmetric subclass of Horndeski theories up to the quadratic order in $\nabla_\mu\nabla_\nu\Phi$ amounts to
\begin{equation}
	A_1=-A_2=2F_X\;, \qquad
	A_3=A_4=A_5=0\;,
\end{equation}
for which $f_1=-f_2=-2XF_X$ and $f_3=f_4=f_5=0$.
This means that some of the EFT coefficients are related to each other and some of them vanish, e.g.,
\begin{equation}
	\bar{K}\alpha+\tilde{\beta}=0\;, \qquad
	\alpha=M_2^2=-M_3^2\;, \qquad
	\zeta={\mathcal M}_1^2={\mathcal M}_2^2={\mathcal M}_3^2=0\;.
\end{equation}
In this case, the expression~\eqref{EFT_coeff_HOST} of the EFT coefficients is consistent with the one in \cite{Mukohyama:2022enj}.

\section{Static and spherically symmetric background}\label{sec:background}

In the last two Sections, we have reviewed a general construction of the EFT and reported a dictionary between the coefficients in the EFT action and the corresponding functions that specify the covariant action of the shift- and $Z_2$-symmetric quadratic higher-order scalar-tensor theories.
In this Section, we consider a static and spherically symmetric background geometry based on the EFT. For simplicity, instead of \eqref{eq:EFT}, we choose the simpler EFT action~\eqref{eq:EFT_dictionary} that includes quadratic higher-order scalar-tensor theories without imposing \eqref{EFT_coeff_HOST} as a starting point.
Also, as mentioned earlier, we focus on black hole solutions and ignore matter fields throughout the present paper.

Let us first specify the background metric which is static and spherically symmetric,
\begin{align}
	ds^2 = -A(r) dt^2 + \frac{dr^2}{B(r)} + r^2 d\Omega^2 \;,
\end{align}
where $d\Omega^2 = d\theta^2 + \sin^2\theta d\phi^2$, and $A(r)$ and $B(r)$ are functions of the areal radius~$r$. Note that, throughout this paper, we do not necessarily impose the condition~$A(r) = B(r)$. The metric above can be brought to the so-called Lema\^{\i}tre coordinates~\cite{Lemaitre:1933gd,Mukohyama:2005rw,Khoury:2020aya,Takahashi:2021bml},
\begin{align}\label{eq:metric_bg}
	ds^2 = -d\tau^2 + [1 - A(r)] d\rho^2+ r^2 d\Omega^2 \;,
\end{align}
with the transformations being
\begin{align}\label{eq:coord_tranf}
	d\tau = dt + \sqrt{\frac{1 - A}{AB}}~dr \;, \qquad 
	d\rho = dt + \frac{dr}{\sqrt{AB(1 - A)}} \;.
\end{align}
Additionally, using the transformations above, we see that the areal radius~$r$ is a function of $\rho-\tau$ and
\begin{align}
	\partial_\rho r = -\dot{r} = \sqrt{\frac{B(1 - A)}{A}} \;, \label{eqn:derivatives_r}
\end{align}
where a dot denotes the derivative with respect to $\tau$.
As is clear in this expression, the $\rho$- and $\tau$-derivatives of $r$ are functions of $r$.

In what follows, for simplicity, we set the EFT coefficient~$f$ in front of the Ricci scalar to be unity and assume $\bar{\Phi}(\tau) = \mu^2\tau$ and $\bar{g}^{\tau\tau}=-1$, so that $\bar{X}=const$.\footnote{Or equivalently, we assume the existence of the Einstein frame which is compatible with $\bar{X}=const$.} Also, we assume the symmetries under $\Phi \rightarrow \Phi + const.$ and $\Phi \rightarrow -\Phi$. With all these assumptions, one can show that the corresponding EFT up to second order on the background~(\ref{eq:metric_bg}) is given by
\begin{align}
	S = \int d^4x \sqrt{-g} \bigg[&\frac{M_\star^2}{2}R - \Lambda(r) - c(r)g^{\tau\tau} - \tilde{\beta}(r) K - \alpha(r)\bar{K}^\mu_\nu K^{\nu}_\mu -\zeta(r) n^\mu\partial_\mu g^{\tau\tau} \nonumber \\ 
	& + \frac{1}{2} m_2^4(r) (\delta g^{\tau\tau})^2 + \frac{1}{2} \tilde{M}_1^3(r) \delta g^{\tau\tau} \delta K + \frac{1}{2} M_2^2(r) \delta K^2 + \frac{1}{2} M_3^2(r) \delta K^\mu_\nu \delta K^\nu_\mu \nonumber \\
	& + \frac{1}{2}\mu_1^2(r) \delta g^{\tau\tau} \delta {}^{(3)}\!R + \frac{1}{2} \lambda_1(r)^\mu_\nu \delta g^{\tau\tau} \delta K^\nu_\mu + \frac{1}{2}{\mathcal M}_1^2(r)(\bar{n}^\mu\partial_\mu\delta g^{\tau\tau})^2 \nonumber \\
	& +\frac{1}{2}{\mathcal M}_2^2(r)\delta K(\bar{n}^\mu\partial_\mu\delta g^{\tau\tau})+\frac{1}{2}{\mathcal M}_3^2(r)\bar{h}^{\mu\nu}\partial_\mu\delta g^{\tau\tau}\partial_\nu\delta g^{\tau\tau} \bigg] \;,
	\label{eq:EFT_HOST}
\end{align}
where the EFT coefficients are now functions of the areal radius~$r=r(\rho-\tau)$ only, respecting the staticity and the spherical symmetry of the background metric. It is now straightforward to write down the background equations of motion. Only the terms in the first line of \eqref{eq:EFT_HOST} contribute to the background equations, and we obtain 
\begin{equation}\label{eq:background}
	M_\star^2\bar{G}_{\mu\nu}=\bar{T}_{\mu\nu} \;,
\end{equation}
where
\begin{align}
	\bar{T}_{\mu\nu}&=
	-(\Lambda-c-\bar{n}^\lambda\partial_\lambda\tilde{\beta}+\alpha\bar{K}^\lambda_\sigma\bar{K}^\sigma_\lambda)\bar{g}_{\mu\nu}+[2c-\bar{n}^\lambda\partial_\lambda\tilde{\beta}+\alpha\bar{K}^\lambda_\sigma\bar{K}^\sigma_\lambda-2\bar{\nabla}_\lambda(\zeta\bar{n}^\lambda)]\bar{n}_\mu\bar{n}_\nu \nonumber \\
	&\quad -2\bar{n}_{(\mu}\partial_{\nu)}\tilde{\beta}+2\alpha\bar{K}_\mu^\lambda\bar{K}_{\lambda\nu}-2\bar{\nabla}_\lambda(\alpha\bar{K}^\lambda_{(\mu}\bar{n}_{\nu)})+\bar{\nabla}_\lambda(\alpha\bar{K}_{\mu\nu}\bar{n}^\lambda)  \;.
\end{align}
Compared with the one in \cite{Mukohyama:2022enj}, we here assume $f=1$ and $\bar{g}^{\tau\tau}=-1$, and there is a contribution from the new tadpole term with $\zeta$ in \eqref{eq:EFT_HOST}. Due to the symmetry of the background, only four out of ten components are independent, which are explicitly written as follows:
\begin{equation}
	\begin{split}
		\Lambda-c
		&=M_\star^2(\bar{G}^\tau{}_\rho-\bar{G}^\rho{}_\rho)\;, \\
		\Lambda+c+\frac{2}{r^2}\sqrt{\frac{B}{A}}\left(r^2\sqrt{1-A}\,\zeta\right)'
		&=-M_\star^2\bar{G}^\tau{}_\tau\;, \\
		\left[\partial_\rho\bar{K}+\frac{1-A}{r}\left(\frac{B}{A}\right)'\,\right]\alpha+\frac{A'B}{2A}\alpha'+\sqrt{\frac{B(1 - A)}{A}}\tilde{\beta}'
		&=-M_\star^2\bar{G}^\tau{}_\rho\;, \\
		\frac{1}{2r^2}\sqrt{\frac{B}{A}}\left[r^4\sqrt{\frac{B}{A}}\left(\frac{1-A}{r^2}\right)'\alpha\right]'
		&=M_\star^2(\bar{G}^\rho{}_\rho-\bar{G}^\theta{}_\theta)\;,
	\end{split} \label{EOM_BG}
\end{equation}
with the relevant components of the (background) Einstein tensor given by
\begin{equation}
	\begin{split}
		\bar{G}^\tau{}_\tau&=-\frac{[r(1-B)]'}{r^2}+\frac{1-A}{r}\left(\frac{B}{A}\right)'\;, \\
		\bar{G}^\tau{}_\rho&=-\frac{1-A}{r}\left(\frac{B}{A}\right)'\;, \\
		\bar{G}^\rho{}_\rho&=-\frac{[r(1-B)]'}{r^2}-\frac{1}{r}\left(\frac{B}{A}\right)'\;, \\
		\bar{G}^\theta{}_\theta&=\frac{B(r^2A')'}{2r^2A}+\frac{(r^2A)'}{4r^2}\left(\frac{B}{A}\right)'\;.
	\end{split}
\end{equation}
Here and in what follows, we use a prime to denote the derivative with respect to $r$. These equations provide relations among the EFT coefficients through given functions~$A(r)$ and $B(r)$.
In particular, one can obtain conditions under which our EFT admits the stealth Schwarzschild(-de Sitter) solutions as an exact solution (see Appendix~\ref{app}).
Conversely, if the EFT coefficients are regarded as an input, then one can fix the functional form of $A(r)$ and $B(r)$ by use of the above equations. For instance, in (quadratic) DHOST theories where $A_1+A_2=0$ [see Eq.~\eqref{HOST}], the third equation in \eqref{EOM_BG} yields $B/A=const$.


\section{Odd-parity perturbations}\label{sec:odd-parity}
In the previous Section, we found the background equations of motion~(\ref{EOM_BG}) derived from the EFT action~(\ref{eq:EFT_HOST}) assuming the shift and $Z_2$ symmetries. In this Section, we analyze the dynamics of linear odd-parity perturbations around the background~(\ref{eq:metric_bg}) based on the quadratic Lagrangian. 
The analysis in this section is a generalization of the one for shift- and $Z_2$-symmetric quadratic DHOST theories performed in \cite{Takahashi:2021bml}, where perturbations about the stealth Schwarzschild-de Sitter solution~\cite{Takahashi:2019oxz,Takahashi:2020hso} were studied.
We show that our results agree with those in \cite{Takahashi:2021bml} when restricted to the stealth solution within shift- and $Z_2$-symmetric DHOST theories.

Let us now analyze the decomposition of metric perturbations. The fact that the metric~(\ref{eq:metric_bg}) admits the $SO(2)$ invariance implies that it is useful to decompose perturbations into the odd and even sectors in the sense of spherical harmonics. Notice that the even and odd sectors are decoupled at the linear level in theories without parity-violating terms.\footnote{For a parity-violating case such as the Chern-Simons theories, see \cite{Motohashi:2011pw} and references therein.} The details of the decomposition into those two sectors for different fields with different spins can be found in \cite{Regge:1957td}. 

We introduce the metric perturbations, $\delta g_{\mu\nu} = g_{\mu\nu} - \bar{g}_{\mu\nu}$, in the odd-parity sector:
\begin{equation}
	\begin{split}
		\delta g_{\tau\tau} &= \delta g_{\tau\rho} = \delta g_{\rho\rho} = 0 \;, \\ 
		\delta g_{\tau a} &= \sum_{\ell,m} r^2 h_{0,\ell m}(\tau,\rho) E_a{}^b \bar{\nabla}_b Y_{\ell m}(\theta,\phi) \;, \\ 
		\delta g_{\rho a} &= \sum_{\ell,m} r^2 h_{1,\ell m}(\tau,\rho) E_a{}^b \bar{\nabla}_b Y_{\ell m}(\theta,\phi) \;, \\
		\delta g_{ab} &= \sum_{\ell,m} r^2 h_{2,\ell m}(\tau,\rho) E_{(a|}{}^c \bar{\nabla}_{c}\bar{\nabla}_{|b)} Y_{\ell m}(\theta,\phi) \;,
	\end{split} \label{eq:pert_h2}
\end{equation}
where $Y_{\ell m}$ is the spherical harmonics, $E_{ab}$ is the completely antisymmetric tensor defined on a 2-sphere, $\bar{\nabla}_a$ refers to the covariant derivative with respect to the metric of the unit 2-sphere, and the indices~$a,b,\cdots$ refer to $\{\theta,\phi\}$. As is well known, the three modes~$h_0$, $h_1$ and $h_2$ are not all physical degrees of freedom due to the residual symmetries in the unitary gauge. Under an infinitesimal odd-parity coordinate transformation, $x^\mu \rightarrow x^\mu + \epsilon^\mu$, with
\begin{align}
	\epsilon^\tau = \epsilon^\rho = 0 \;, \quad \epsilon^a = \sum_{\ell,m} \Xi_{\ell m} (\tau,\rho) E^{ab} \bar{\nabla}_b Y_{\ell m}(\theta,\phi) \;,
\end{align}
the unitary gauge is preserved and the components~$h_0$, $h_1$, and $h_2$ transform as
\begin{align}
	h_0 \rightarrow h_0 - \dot{\Xi} \;, \quad h_1 \rightarrow h_1 - \partial_\rho\Xi \;, \quad h_2 \rightarrow h_2 - 2\Xi \label{eq:gauge_h0_h1} \;.
\end{align}
One can simplify the analysis by setting $m = 0$. In fact, this simplification can be justified by the spherical symmetry of the background. Therefore, it is more convenient to work with the Legendre polynomials~$P_\ell(\cos\theta)$ instead of the spherical harmonics. In what follows, we use $h_0$, $h_1$, and $h_2$ to denote the coefficients of $P_\ell(\cos\theta)$.

For general multipoles~$\ell \geq 2$, one can fix $h_2\to 0$ by choosing $\Xi = h_2/2$, which is a complete gauge fixing and can be done at the Lagrangian level.
In Subsection~\ref{ssec:multipole}, we will see that there is only one dynamical degree of freedom in the odd-parity sector, which corresponds to one of the two helicity modes of gravitational waves.
On the other hand, the discussion here does not apply to dipole perturbations, which we shall discuss separately in Subsection~\ref{ssec:dipole}.

\subsection{\texorpdfstring{Odd-parity perturbations with $\ell \geq 2$}{General multipoles}}\label{ssec:multipole}
Let us now turn to the EFT action relevant to the odd-parity perturbations. By construction, the background metric~(\ref{eq:metric_bg}) is even under parity and so are the quantities evaluated on the background. Besides, the operators~$\delta g^{\tau\tau}$ and $\delta K$ are even so that the terms in (\ref{eq:EFT_HOST}) that contain them may be safely omitted. Therefore, with all the considerations above, the EFT action we are going to consider for odd-parity perturbations is 
\begin{align}
	S_{\rm odd} = \int d^4x \sqrt{-g} \bigg[\frac{M_\star^2}{2}R - \Lambda(r) - c(r)g^{\tau\tau} -\tilde{\beta}(r) K - \alpha(r)\bar{K}^{\mu}_\nu K^\nu_{\mu} + \frac{1}{2} M_3^2(r) \delta K^\mu_\nu \delta K^\nu_\mu \bigg] \;. \label{eq:EFT_odd}
\end{align}
Note that one advantage of considering the odd-parity sector is that one does not need to take into account the Stueckelberg field~$\pi$ since it is only present in the even-parity sector.
Hence, working in the unitary gauge is sufficient to fully describe the dynamics of the odd-parity perturbations.
Related to this point, a possible extra degree of freedom due to higher derivatives of the scalar field does not show up in the odd sector.
As a result, the analysis of odd-parity perturbations in the present paper applies to scalar-tensor theories with non-degenerate higher-derivative terms.
Likewise, it also applies to theories with a non-dynamical scalar field, e.g., the cuscuton~\cite{Afshordi:2006ad} or its extension~\cite{Iyonaga:2018vnu,Iyonaga:2020bmm}.

From the action~(\ref{eq:EFT_odd}), the quadratic action for $h_0$ and $h_1$ after integrating over the angular variables and a few integration by parts is given by $S_2 = \int d\tau d\rho~\mathcal{L}_2$, with
\begin{align}
	\frac{2\ell + 1}{2 \pi j^2}\mathcal{L}_2 = p_1 h_0^2 + p_2 h_1^2 + p_3 [(\dot{h}_1 - \partial_\rho h_0)^2 + 2p_4 h_1 \partial_\rho h_0] \;, \label{L2_odd_Ein}
\end{align}
with the coefficients~$p$'s defined as\footnote{Since $p_3$ and $p_4$ are now functions of $r$ only, one can express the last term in $p_2$ in terms of $r$-derivative by use of \eqref{eqn:derivatives_r}. Nevertheless, we keep the $\tau$-derivative for notational simplicity.}
\begin{equation}
	\begin{split}
		p_1 &\equiv  \frac{1}{2}(j^2-2)r^2\sqrt{1-A}\,(M_\star^2 + M_3^2) \;, \quad
		p_2 \equiv -(j^2-2)\frac{r^2M_\star^2}{2\sqrt{1-A}} + (p_3p_4)^{\boldsymbol{\cdot}} \;, \\
		p_3 &\equiv \frac{(M^2_\star + M_3^2) r^4}{2\sqrt{1 - A}} \;, \quad
		p_4 \equiv \sqrt{\frac{B}{A(1 - A)}}\left(\frac{A'}{2}+\frac{1-A}{r}\right) \frac{\alpha + M_3^2}{M_\star^2 + M_3^2} \;.
	\end{split} \label{eq:odd_2_ein}
\end{equation}
Here, $j^2 \equiv \ell(\ell + 1)$ and we have used the background equations of motion~\eqref{EOM_BG} to remove the tadpole functions~$\Lambda$, $c$, and $\tilde{\beta}$. 
One can in principle neglect the derivatives of $M_3^2$ and $\alpha$ since they are typically suppressed by the energy scale~$E$ [see Eq.~(\ref{eq:energy_E})] that is assumed to be much lower than the cutoff scale of the EFT we are interested in.\footnote{Furthermore, within the class of shift- and $Z_2$-symmetric quadratic higher-order scalar-tensor theories~\eqref{HOST}, the parameters~$\alpha$ and $M_3^2$ are just functions of $\bar{X}$ [see Eq.~(\ref{EFT_coeff_HOST})] that is constant in our setup, so that all derivatives acting on those parameters trivially vanish.} Nevertheless, in order to accommodate more general situations, we keep all of them in our expressions.

Before going to further analysis, let us now comment on the coefficients~\eqref{eq:odd_2_ein}. 
First, there is one cross term with the coefficient~$p_4$. We will see below that there is no obstacle of showing that there is only one physical degree of freedom even with the presence of this cross term. 
The coefficient~$p_4$ vanishes when one imposes $\alpha + M_3^2 = 0$, which can be realized in, for instance, the shift- and $Z_2$-symmetric quadratic higher-order scalar-tensor theories [see Eq.~(\ref{EFT_coeff_HOST})].
For now, however, we consider $p_4$ as a function of $r$.
Furthermore, it is interesting to point out that the above expressions of $p_1$, $p_2$, and $p_3$ are consistent with the ones in \cite{Takahashi:2021bml} where the stealth Schwarzschild-de Sitter solutions in the context of shift- and $Z_2$-symmetric quadratic DHOST theories were studied.

Let us now get back to the Lagrangian~(\ref{L2_odd_Ein}). Following the procedure introduced in \cite{DeFelice:2011ka}, we perform an integration by parts and complete the square of terms containing derivatives to obtain
\begin{align}
	\frac{2\ell + 1}{2 \pi j^2}\mathcal{L}_2 = p_1 h_0^2 + \tilde{p}_2 h_1^2 + p_3 (\dot{h}_1 - \partial_\rho h_0 - p_4h_1)^2  \;, \label{eq:L_odd_2}
\end{align}
where we have defined
\begin{equation}
	\tilde{p}_2\equiv p_2 -(p_3 p_4)^{\boldsymbol{\cdot}} - p_3 p_4^2\;.
\end{equation}
We then introduce an auxiliary variable~$\chi$ (whose mass dimension is 2) such that the Lagrangian above can be rewritten as 
\begin{align}
	\frac{2\ell + 1}{2 \pi j^2}\mathcal{L}_2 = p_1 h_0^2 + \tilde{p}_2 h_1^2
	+ p_3[-\chi^2 + 2\chi (\dot{h}_1 - \partial_\rho h_0 - p_4h_1)]  \;. \label{eq:L_2_chi}
\end{align}
Notice that the Lagrangian (\ref{eq:L_odd_2}) can be recovered from \eqref{eq:L_2_chi} by integrating out $\chi$. 
Indeed, the equation of motion for $\chi$ yields $\chi = \dot{h}_1 - \partial_\rho h_0 - p_4h_1$, which can be substituted back into \eqref{eq:L_2_chi} to recover the Lagrangian~\eqref{eq:L_odd_2}. Then, the Euler-Lagrange equations for $h_0$ and $h_1$ obtained from (\ref{eq:L_2_chi}) can be algebraically solved for $h_0$ and $h_1$ as
\begin{align}
	h_0 &= - \frac{\partial_\rho(p_3 \chi) }{p_1} \;, \qquad
	h_1 = \frac{(p_3 \chi)^{\boldsymbol{\cdot}} + p_3 p_4 \chi }{\tilde{p}_2} \;. \label{eq:solh0h1}
\end{align}
Plugging these solutions back into (\ref{eq:L_2_chi}), we obtain the quadratic Lagrangian for the field~$\chi$ only,
\begin{align}\label{eq:chi_s}
	\frac{(j^2 - 2)(2\ell + 1)}{2 \pi j^2} \mathcal{L}_2 = s_1 \dot{\chi}^2 - s_2 (\partial_\rho\chi)^2 - s_3 \chi^2 \;,
\end{align}
where the parameters~$s_1$--$s_3$ are given by
\begin{equation}
	\begin{split}
		s_1 &= -\frac{(j^2 - 2)p_3^2}{\tilde{p}_2} = \frac{j^2 - 2}{2\sqrt{1 - A}}\frac{(M_\star^2 + M_3^2)^{2}r^6}{(j^2 - 2) M_\star^2 + (M_\star^2  + M_3^2) r^2 p_4^2} \;, \\
		s_2 &= \frac{(j^2 - 2)p_3^2}{ p_1}  = \frac{(M_\star^2 + M_3^2)r^6}{2 (1 - A)^{3/2}} \;, \\
		s_3 &= (j^2-2)p_3\left[1+\frac{p_3p_4^2}{\tilde{p}_2}-\left(\frac{p_1+\tilde{p}_2}{p_1\tilde{p}_2}\dot{p}_3\right)^{\boldsymbol{\cdot}}-p_3\left(\frac{p_4}{\tilde{p}_2}\right)^{\boldsymbol{\cdot}}\,\right] = j^2 \frac{(M^2_\star + M_3^2) r^4}{2\sqrt{1 - A}} + {\cal O}(j^0)\;.
	\end{split}
\end{equation}

From the Lagrangian~\eqref{eq:chi_s}, one can read off squared sound speeds as follows.
The squared sound speed~$c_\rho^2$ along the $\rho$-direction can be defined by
\begin{equation}
	c_\rho^2 = \frac{\bar{g}_{\rho\rho}}{|\bar{g}_{\tau\tau}|}\frac{s_2}{s_1} =  \frac{M_\star^2}{M_\star^2 + M_3^2} + \frac{r^2 p_4^2}{j^2 - 2} \;, \label{eq:c2_rho}
\end{equation}
where we have inserted the factor~$\bar{g}_{\rho\rho}/|\bar{g}_{\tau\tau}|$ so that the sound speed indeed represents $\delta(\text{proper distance})$ divided by $\delta\text(\mbox{proper time})$.
Hence, so long as $p_4\ne 0$ (i.e., $\alpha+M_3^2\ne 0$ in terms of EFT coefficients), $c_\rho^2$ depends on the multipole index of the mode functions.
It should be noted that, in the case where $p_4$ is non-vanishing, one requires that $r p_4$ remains finite at large $r$ in order to avoid a diverging sound speed.\footnote{This is indeed the case for the Schwarzschild background if $\alpha$ and $M_3^2$ are constant.}
Therefore, if $p_4$ is non-vanishing then it must be a function of $r$, which asymptotically goes to zero at least as fast as $1/r$. However, as we will see in Section~\ref{ssec:dipole}, this non-constant $p_4$ case does not allow for a slowly rotating black hole solution [see the discussion below Eq.~(\ref{eq:C(tau)})].

The squared sound speed~$c_\theta^2$ along angular directions can be defined by
\begin{equation}
	c_\theta^2 = \lim_{\ell\to\infty}\frac{r^2}{|\bar{g}_{\tau\tau}|}\frac{s_3}{j^2s_1}
	=\frac{M_\star^2}{M_\star^2 + M_3^2}\;, \label{eq:c2_theta}
\end{equation}
where the factor~$r^2/|\bar{g}_{\tau\tau}|$ has been inserted for the same reason as $\bar{g}_{\rho\rho}/|\bar{g}_{\tau\tau}|$ in \eqref{eq:c2_rho}.
Note that $c_\rho^2\ne c_\theta^2$ in general, and the difference is characterized by $p_4$ (i.e., $\alpha+M_3^2$).
Therefore, in shift- and reflection-symmetric quadratic higher-order scalar-tensor theories where $\alpha+M_3^2=0$ [see Eq.~\eqref{EFT_coeff_HOST}], the two sound speeds coincide:
\begin{align}\label{eq:c_T_special}
	c_\rho^2 = c_\theta^2 = \frac{M_\star^2 }{M_\star^2 + M_3^2} \equiv c_T^2 \;,
\end{align}
which is consistent with the result of \cite{Takahashi:2021bml}. Moreover, in this particular case, we see that the difference between the sound speed $c_T^2$ and unity is determined by the operator~$M_3^2$ of the EFT. The same expression of $c_T^2$ for cosmological backgrounds in the EFT of dark energy can be found, for example, in \cite{Creminelli:2017sry}. 

We require the absence of ghost and gradient instabilities,
\begin{equation}
	s_1>0\;, \qquad
	c_\rho^2>0\;, \qquad
	c_\theta^2>0\;,
\end{equation}
which can be realized if our EFT parameters satisfy
\begin{align}
	M_\star^2 + M_3^2 > 0 \;, \qquad
	M_\star^2  > 0 \;.
\end{align}
These conditions are again consistent with those in \cite{Takahashi:2021bml}. Notice that the conditions above hold only at the linear level, whereas the stability conditions become more complicated when non-linearities are taken into account.\footnote{One example where the perturbations on a cosmological background become pathological was studied in \cite{Creminelli:2019kjy}. The instabilities were found in a non-linear regime in the presence of gravitational wave background, taking into account the cubic and quartic Galilean operators.} 

Let us now derive the generalized Regge-Wheeler equation from the Lagrangian~(\ref{eq:chi_s}). Since one usually derives the Regge-Wheeler equation in the Schwarzschild coordinates~$\{t, r, \theta, \phi\}$, let us perform the inverse transformation of \eqref{eq:coord_tranf} by use of
\begin{align}
	dt = \frac{1}{A} d\tau - \frac{1 - A}{A} d\rho \;, \qquad
	dr = -\sqrt{\frac{B(1 - A)}{A}} d\tau + \sqrt{\frac{B(1 - A)}{A}} d\rho \;. \label{Lemaitre_to_Sch}
\end{align}
Then, the Lagrangian~(\ref{eq:chi_s}) becomes
\begin{align}\label{eq:L2_odd_a}
	\frac{(j^2 - 2)(2\ell + 1)}{2 \pi j^2} \mathcal{L}_2 = a_1 (\partial_t \chi)^2 - a_2 (\partial_r \chi)^2 + 2a_3 (\partial_t \chi) (\partial_r \chi) - a_4\chi^2 \;,
\end{align}
with
\begin{equation}
    \begin{split}
	&a_1=\frac{s_1-(1-A)^2s_2}{\sqrt{A^3B(1-A)}}\;, \qquad
	a_2=\sqrt{\frac{B(1-A)}{A}}(s_2-s_1)\;, \\
	&a_3=\frac{(1-A)s_2-s_1}{A}\;, \qquad
	a_4=\sqrt{\frac{A}{B(1-A)}}s_3\;.
	\end{split}
\end{equation}
Note that we have multiplied the Lagrangian by the following Jacobian determinant associated with the coordinate transformation~\eqref{Lemaitre_to_Sch}:
\begin{equation}
	\left|\frac{\partial(\tau,\rho)}{\partial(t,r)}\right|
	= \sqrt{\frac{A}{B(1-A)}}\;.
\end{equation}
Written in terms of the Schwarzschild coordinates, all the coefficients defined above are independent of $t$. The Lagrangian~\eqref{eq:L2_odd_a} contains the cross term~$(\partial_t \chi) (\partial_r \chi)$, and thus the ``sound cone'' of odd-parity gravitational waves is tilted relative to the timelike Killing vector~$(\partial/\partial t)^{\mu}$, reflecting the fact that $\bar{g}^{\mu\nu}\partial_{\nu}\bar{\Phi}$ is not proportional to the timelike Killing vector. As explained in \cite{Takahashi:2019oxz}, the cross term in (\ref{eq:L2_odd_a}) can be removed by performing a coordinate transformation,
\begin{align}
	\tilde{t} = t + \int \frac{a_3}{a_2} dr \;.
\end{align}
Note that the radial coordinate remains the same. Therefore, we have
\begin{align}
	\frac{(j^2 - 2)(2\ell + 1)}{2 \pi j^2} \mathcal{L}_2  = \tilde{a}_1 (\partial_{\tilde{t}} \chi)^2 - a_2 (\partial_r \chi)^2 - a_4\chi^2 \;,
\end{align}
where now the coefficient~$\tilde{a}_1$ is given by
\begin{align}
	\tilde{a}_1 = a_1 + \frac{a_3^2}{a_2} \;.
\end{align}

We now introduce the tortoise coordinate,
\begin{align}
	r_* = \int \frac{1}{\sqrt{AB}} dr \;,
\end{align}
and a new variable,
\begin{align}
	\Psi = \sqrt{\Gamma}\,\chi \;, \qquad
	\Gamma \equiv \frac{a_2}{\sqrt{AB}} \;.
\end{align}
With the definitions above, the equation of motion for $\Psi$ reads
\begin{align}\label{eq:RG}
	\frac{\partial^2 \Psi}{\partial \tilde{t}^2} - c_{r_*}^2 \frac{\partial^2 \Psi}{\partial r_*^2} + V_{\rm eff} \Psi = 0 \;, 
\end{align}
where we have defined $c_{r_*}^2 \equiv a_2/(AB\tilde{a}_1)$ and the effective potential,
\begin{align}
	V_{\rm eff} \equiv \frac{a_4}{\tilde{a}_1} + \frac{1}{2\sqrt{AB}\,\tilde{a}_1}\frac{d^2 \Gamma}{dr_*^2} - \frac{1}{4\tilde{a}_1 a_2}\bigg(\frac{d \Gamma}{d r_*}\bigg)^2 \;.
\end{align}
Solving (\ref{eq:RG}) under appropriate boundary conditions at the sound horizon for the odd modes and at spatial infinity gives rise to the spectrum of quasinormal modes. Note that a similar analysis of the Regge-Wheeler equation of black hole perturbations around a stealth Schwarzschild solution can be found in \cite{Langlois:2021aji,Langlois:2022ulw}. 

\subsection{\texorpdfstring{Dipole $\ell = 1$}{Dipole}}\label{ssec:dipole}
In this Subsection, we study the dynamics of dipole perturbations in the odd-parity sector, which is related to the slow rotation of black holes. Notice that, in this case, the perturbation~$h_2$ in \eqref{eq:pert_h2} intrinsically vanishes simply due to the fact that $h_2$ was defined with a second derivative acting on $P_\ell(\cos\theta)$ that is zero when $\ell = 1$. Therefore, we still have a gauge freedom to remove either $h_0$ or $h_1$. 

In principle, looking at (\ref{eq:gauge_h0_h1}), one could simply set $\partial_\rho \Xi = h_1$ such that $h_1\to 0$; however, this gauge fixing is incomplete in the sense that it only fixes the $\rho$-derivative of the gauge parameter~$\Xi$, which means that one still has a freedom to change $\Xi$ with an arbitrary function of $\tau$. 
As a result, we will see explicitly below that imposing directly the gauge condition~$h_1 = 0$ in the Lagrangian~(\ref{L2_odd_Ein}) leads to a loss of an independent equation of motion and hence an incorrect result. (See \cite{Motohashi:2016prk} for a more detailed discussion on this point.)
Hence, we should impose such an incomplete gauge fixing after deriving the equations of motion for $h_0$ and $h_1$.

For $\ell=1$, the Lagrangian~\eqref{L2_odd_Ein} takes the form
\begin{align}
	\frac{3}{4 \pi}\mathcal{L}_2 = (p_3p_4)^{\boldsymbol{\cdot}} h_1^2 + p_3 [(\dot{h}_1 - \partial_\rho h_0)^2 + 2p_4 h_1 \partial_\rho h_0] \;,
\end{align}
from which the equations of motion for $h_0$ and $h_1$ can be derived respectively as
\begin{equation}
	\begin{split}
		\partial_\rho[p_3(\dot{h}_1 - \partial_\rho h_0)] + \partial_\rho(p_3 p_4 h_1) &= 0 \;, \\ 
		[p_3 (\dot{h}_1 - \partial_\rho h_0)]^{\boldsymbol{\cdot}} - (p_3p_4)^{\boldsymbol{\cdot}} h_1 - p_4 p_3 \partial_\rho h_0 &= 0 \;.
	\end{split}
\end{equation}
Setting $h_1 = 0$, we have
\begin{equation}
	\partial_\rho(p_3\partial_\rho h_0)=0\;, \qquad
	(p_3\partial_\rho h_0)^{\boldsymbol{\cdot}}+p_4 p_3 \partial_\rho h_0=0\;.
\end{equation}
The first equation yields $p_3 \partial_\rho h_0 = C_1(\tau)$ with $C_1(\tau)$ being an arbitrary function of $\tau$.
Plugging this into the second equation, we find
\begin{equation}
	\dot{C}_1(\tau)+p_4C_1(\tau)=0\;. \label{eq:C(tau)}
\end{equation}
We recall that $p_4$ is a function of $r=r(\rho-\tau)$ in general [see Eq.~\eqref{eq:odd_2_ein}].
Actually, as we discussed below \eqref{eq:c2_rho}, in order for the sound speed in the radial direction to be finite at large $r$, $p_4$ must be a function of $r$ which decays at least as fast as $1/r$.
If $p_4$ is non-vanishing and has a non-trivial $r$-dependence, then the only solution to \eqref{eq:C(tau)} is a trivial one, $C_1(\tau)=0$, which suggests that a slowly rotating black hole solution is prohibited in this case.\footnote{A similar situation happens in the Einstein-aether theory with a hypersurface-orthogonal aether configuration~\cite{Barausse:2012ny}. On the other hand, a rotating solution exists in the infrared limit of non-projectable Ho\v{r}ava-Lifshitz gravity, which is equivalent to the Einstein-aether theory with the hypersurface orthogonality condition imposed before the variation~\cite{Wang:2012at,Barausse:2012qh,Wang:2012nv}. Interestingly, the infrared limit of Ho\v{r}ava-Lifshitz gravity is included in the action~\eqref{HOST2} for shift- and reflection-symmetric quadratic higher-order scalar-tensor theories. Hence, our results in this Subsection are consistent with those in \cite{Wang:2012at,Barausse:2012qh,Wang:2012nv}.}
Hereafter, we therefore assume that $p_4=0$, i.e., $\alpha+M_3^2=0$ in terms of EFT coefficients. Note that the condition~$\alpha+M_3^2=0$ is automatically satisfied by shift- and reflection-symmetric quadratic higher-order scalar-tensor theories [see Eq.~\eqref{EFT_coeff_HOST}]. In this case, \eqref{eq:C(tau)} implies that $C_1$ is constant, and hence
\begin{equation}
	h_0=C_1 \int \frac{d\rho}{p_3}
	=2C_1 \int d\rho\,\frac{\partial_\rho r}{(M^2_\star + M_3^2) r^4}\sqrt{\frac{A}{B}}\;.
\end{equation}
This indicates that the explicit functional form of $h_0$ depends on the background functions~$A(r)$ and $B(r)$ as well as the EFT coefficient~$M_3^2(r)$.
If $M_3^2$ is constant and $A=B$, then the integral can be performed analytically to yield
\begin{align}\label{h0_odd_AB_equal}
	h_0 = - \frac{J}{4\pi (M_\star^2 + M_3^2 ) r^3} \;, \quad J\equiv \frac{8\pi C_1}{3}\;,
\end{align}
where we have used the residual gauge degree of freedom~$\Xi=\Xi(\tau)$ to fix the ($\tau$-dependent) integration constant.
Interestingly, \eqref{h0_odd_AB_equal} has the same form as the ($t\phi$)-component of the Kerr metric expanded up to the first order in the angular momentum~$J$.
This result is consistent with the ones found in \cite{Takahashi:2019oxz,Takahashi:2021bml}.
On the other hand, if $A \neq B$ and/or $M_3^2\ne const$, the solution for $h_0$ is not of the form~\eqref{h0_odd_AB_equal}, implying that a rotating black hole solution does not belong to the Kerr family in general, even at the linear level.
We leave this investigation to future work.


\section{Conclusions} \label{sec:conclusions}
Recently, the Effective Field Theory (EFT) of black hole perturbations in the context of scalar-tensor theories with a timelike scalar profile was formulated in \cite{Mukohyama:2022enj}. The resulting EFT can in fact be applied to an arbitrary spacetime geometry with or without black holes as far as the gradient of the scalar field is timelike. It was shown that imposing a set of consistency relations associated to the spatial coordinates guarantees that the EFT action in the unitary gauge is invariant under the 3d diffeomorphism. 

In Section~\ref{sec:overview} of the present paper, we reviewed the construction of the EFT in the unitary gauge. We also took into account the terms that contain derivatives acting on $\delta g^{\tau\tau}$ to accommodate the shift- and $Z_2$-symmetric quadratic higher-order scalar-tensor theories. We then obtained the corresponding consistency relations taking into account those additional terms in the EFT. In Section~\ref{sec:dictionary}, we restricted our considerations to the shift- and $Z_2$-symmetric quadratic higher-order scalar-tensor theories with $\bar{X} = const.$ and $\bar{g}^{\tau\tau} = -1$ and found the dictionary between our EFT parameters and the functions of such covariant theories.

The background analysis of the EFT including the extra terms due to the shift- and $Z_2$-symmetric quadratic higher-order scalar-tensor theories was performed in Section~\ref{sec:background}. There, we assumed the background metric to be static and spherically symmetric. We thus obtained the background equations of motion which allow us to express some coefficients in the EFT action in terms of the background functions~$A(r)$ and $B(r)$.

Lastly, in Section~\ref{sec:odd-parity}, we analyzed the dynamics of odd-parity perturbations at the second order. We found that the sound speed in the radial direction has a non-trivial dependence on the multipole index~$\ell$ in general. In terms of EFT coefficients, the $\ell$-dependent part is controlled by the quantity~$\alpha+M_3^2$. Interestingly, a non-vanishing $\alpha+M_3^2$ also results in a deviation between the sound speeds in the radial and angular directions. We checked that, once restricted to the stealth Schwarzschild-de Sitter solution in shift- and $Z_2$-symmetric DHOST theories where $\alpha+M_3^2=0$, both the sound speeds turn out to be the same as those in \cite{Takahashi:2021bml}. We also derived the generalized Regge-Wheeler equation in the odd sector. Finally, we studied the odd-parity perturbations with $\ell = 1$ which are related to the slow rotation of a black hole.

There are several directions we would like to explore in future work. First, an obvious application of our current work is to determine the spectrum of quasinormal modes in the odd-parity sector using the generalized Regge-Wheeler equation we found in Section~\ref{sec:odd-parity}. This would be a first step for our EFT to make a connection with observations. Second, it would be nice to use our EFT to compute tidal Love numbers of black holes. It was shown that the Love number of Schwarzschild and Kerr black holes in GR vanishes, see e.g.~\cite{Hui:2020xxx,Charalambous:2021mea}, but it is generically non-zero in some classes of modified gravity theories~\cite{Bernard:2019yfz}. Also, it is worth investigating the dynamics of even-parity perturbations including the spectrum of quasinormal modes and tidal Love numbers. In the even-parity sector, we expect that the strong coupling problem of perturbations around stealth solutions (see e.g.~\cite{deRham:2019gha,Motohashi:2019ymr,Gorji:2020bfl,Takahashi:2021bml,Gorji:2021isn}) is not present in general due to the fact that the so-called scordatura effect~\cite{Motohashi:2019ymr}\footnote{Such an effect was already known in the context of ghost condensation~\cite{Arkani-Hamed:2003pdi,Mukohyama:2005rw}.} is already implemented in our EFT action. Additionally, the procedure for constructing the EFT explained briefly in Section~\ref{sec:overview} can be used to formulate the EFT of vector-tensor theories on arbitrary background metric. This also serves as a generalization of the EFT of vector-tensor theories on a cosmological background developed in \cite{Aoki:2021wew}. We leave this to future work~\cite{Aoki:2022_prep}. Finally, it is theoretically and phenomenologically interesting to formulate the EFT of perturbations on a rotating black hole background accompanied by a timelike scalar profile.\footnote{See \cite{Hui:2021cpm} for a formulation of the EFT on a slowly rotating black hole background with a spacelike (instead of timelike) scalar profile.} This would shed light on how the dynamics of perturbations gets modified due to the presence of angular momentum of a black hole.

\vskip5mm
{\bf Note added}: While we were finalizing this paper, we were informed of a research by Justin Khoury, Toshifumi Noumi, Mark Trodden, and Sam S.~C.~Wong on a similar subject~\cite{Khoury:2022zor}.
Their results are consistent with ours on overlapping parts.
We would like to thank them for their kind correspondence.

\section*{Acknowledgements}
This work was supported by World Premier International Research Center Initiative (WPI), MEXT, Japan. The work of S.~M.\ was supported by JSPS KAKENHI No.\ 17H02890, No.\ 17H06359. 
The work of K.~T.\ was supported by JSPS KAKENHI Grant No.\ JP21J00695.
The work of V.~Y.\ was supported by JSPS KAKENHI Grant No.\ JP22K20367.


\appendix

\section{Existence conditions for stealth solutions}\label{app}

Let us study under which conditions our EFT admits the stealth Schwarzschild(-de Sitter) solution as an exact solution.
The conditions are obtained by plugging the functions
\begin{equation}
	A=B=1-\frac{r_s}{r}-\frac{\Lambda_{\rm eff}}{3}r^2\;,
\end{equation}
into the background equations of motion~\eqref{EOM_BG}.
Here, $r_s$ and $\Lambda_{\rm eff}$ are constants, and the background Einstein tensor is given by $\bar{G}^\mu{}_\nu=-\Lambda_{\rm eff}\delta^\mu{}_\nu$.
A similar strategy was employed to clarify the existence conditions for stealth solutions, e.g., in \cite{Motohashi:2018wdq,Minamitsuji:2018vuw,Motohashi:2019sen,Takahashi:2020hso}.
After some simplifications, we have the following conditions on tadpole functions:
\begin{equation}
	\Lambda-c=M_\star^2 \Lambda_{\rm eff}\;, \qquad
	c+\frac{1}{r^2}\left(r^2\sqrt{1-A}\,\zeta\right)'=0\;, \qquad
	\alpha'=0\;, \qquad
	\bar{K}'\alpha+\tilde{\beta}'=0\;.
	\label{eq:existence_stealth}
\end{equation}

In the case of shift- and reflection-symmetric higher-order scalar-tensor theories, the last condition in \eqref{eq:existence_stealth} cannot be satisfied in general.
Indeed, by use of the dictionary~\eqref{EFT_coeff_HOST} as well as \eqref{eq:fs}, we have
\begin{equation}
	\bar{K}'\alpha+\tilde{\beta}'
	=-2\bar{K}'(\bar{f}_1+\bar{f}_2)
	=2\bar{X}\bar{K}'(\bar{A}_1+\bar{A}_2)\;.
\end{equation}
Here, $\bar{A}_1$ and $\bar{A}_2$ are the coefficient functions~$A_1(X)$ and $A_2(X)$ in the action~\eqref{HOST} evaluated at the background, $X=\bar{X}$.
This means that theories with $A_1+A_2\ne 0$ are incompatible with the set of conditions~\eqref{eq:existence_stealth} and hence do not accommodate the stealth Schwarzschild(-de Sitter) solution as an exact solution.
In particular, U-DHOST theories, for which $A_1+A_2$ can be non-vanishing, are incompatible with the condition~$\bar{K}'\alpha+\tilde{\beta}'=0$ in general.
For DHOST theories, the last two conditions in \eqref{eq:existence_stealth} are satisfied by default, and the first two conditions just fix the parameters of the solution~$\Lambda_{\rm eff}$ and $\bar{X}$ in terms of the coefficient functions in the action~\eqref{HOST}, which is consistent with the result of \cite{Takahashi:2019oxz,Takahashi:2020hso}.

{}
\bibliographystyle{utphys}
\bibliography{bib_v4}

\providecommand{\href}[2]{#2}\begingroup\raggedright\begin{thebibliography}{10}

\bibitem{jordan1955schwerkraft}
P.~Jordan, {\em Schwerkraft und Weltall: Grundlagen d. theoret. Kosmologie. Mit
  13 Abb}.
\newblock Die Wissenschaft. Vieweg, 1955.

\bibitem{Brans:1961sx}
C.~Brans and R.~H. Dicke, ``{Mach's principle and a relativistic theory of
  gravitation},'' {\em Phys. Rev.} {\bf 124} (1961) 925--935.

\bibitem{Horndeski:1974wa}
G.~W. Horndeski, ``{Second-order scalar-tensor field equations in a
  four-dimensional space},'' {\em Int.J.Theor.Phys.} {\bf 10} (1974) 363--384.

\bibitem{Deffayet:2011gz}
C.~Deffayet, X.~Gao, D.~A. Steer, and G.~Zahariade, ``{From k-essence to
  generalised Galileons},'' {\em Phys. Rev. D} {\bf 84} (2011) 064039,
  \href{https://arxiv.org/abs/1103.3260}{{\tt 1103.3260}}.

\bibitem{Kobayashi:2011nu}
T.~Kobayashi, M.~Yamaguchi, and J.~Yokoyama, ``{Generalized G-inflation:
  Inflation with the most general second-order field equations},'' {\em
  Prog.Theor.Phys.} {\bf 126} (2011) 511--529,
  \href{https://arxiv.org/abs/1105.5723}{{\tt 1105.5723}}.

\bibitem{Langlois:2015cwa}
D.~Langlois and K.~Noui, ``{Degenerate higher derivative theories beyond
  Horndeski: evading the Ostrogradski instability},'' {\em JCAP} {\bf 02}
  (2016) 034, \href{https://arxiv.org/abs/1510.06930}{{\tt 1510.06930}}.

\bibitem{Crisostomi:2016czh}
M.~Crisostomi, K.~Koyama, and G.~Tasinato, ``{Extended Scalar-Tensor Theories
  of Gravity},'' {\em JCAP} {\bf 04} (2016) 044,
  \href{https://arxiv.org/abs/1602.03119}{{\tt 1602.03119}}.

\bibitem{BenAchour:2016fzp}
J.~Ben~Achour, M.~Crisostomi, K.~Koyama, D.~Langlois, K.~Noui, and G.~Tasinato,
  ``{Degenerate higher order scalar-tensor theories beyond Horndeski up to
  cubic order},'' {\em JHEP} {\bf 12} (2016) 100,
  \href{https://arxiv.org/abs/1608.08135}{{\tt 1608.08135}}.

\bibitem{Langlois:2018dxi}
D.~Langlois, ``{Dark energy and modified gravity in degenerate higher-order
  scalar\textendash{}tensor (DHOST) theories: A review},'' {\em Int. J. Mod.
  Phys. D} {\bf 28} (2019), no.~05 1942006,
  \href{https://arxiv.org/abs/1811.06271}{{\tt 1811.06271}}.

\bibitem{Kobayashi:2019hrl}
T.~Kobayashi, ``{Horndeski theory and beyond: a review},'' {\em Rept. Prog.
  Phys.} {\bf 82} (2019), no.~8 086901,
  \href{https://arxiv.org/abs/1901.07183}{{\tt 1901.07183}}.

\bibitem{DeFelice:2018ewo}
A.~De~Felice, D.~Langlois, S.~Mukohyama, K.~Noui, and A.~Wang, ``{Generalized
  instantaneous modes in higher-order scalar-tensor theories},'' {\em Phys.
  Rev. D} {\bf 98} (2018), no.~8 084024,
  \href{https://arxiv.org/abs/1803.06241}{{\tt 1803.06241}}.

\bibitem{DeFelice:2021hps}
A.~De~Felice, S.~Mukohyama, and K.~Takahashi, ``{Nonlinear definition of the
  shadowy mode in higher-order scalar-tensor theories},'' {\em JCAP} {\bf 12}
  (2021), no.~12 020, \href{https://arxiv.org/abs/2110.03194}{{\tt
  2110.03194}}.

\bibitem{DeFelice:2022xvq}
A.~De~Felice, S.~Mukohyama, and K.~Takahashi, ``{Built-in scordatura in
  U-DHOST},'' {\em Phys. Rev. Lett.} {\bf 129} (2022) 031103,
  \href{https://arxiv.org/abs/2204.02032}{{\tt 2204.02032}}.

\bibitem{Bekenstein:1992pj}
J.~D. Bekenstein, ``{The Relation between physical and gravitational
  geometry},'' {\em Phys. Rev. D} {\bf 48} (1993) 3641--3647,
  \href{https://arxiv.org/abs/gr-qc/9211017}{{\tt gr-qc/9211017}}.

\bibitem{Bruneton:2007si}
J.-P. Bruneton and G.~Esposito-Far{\`e}se, ``{Field-theoretical formulations of
  MOND-like gravity},'' {\em Phys. Rev. D} {\bf 76} (2007) 124012,
  \href{https://arxiv.org/abs/0705.4043}{{\tt 0705.4043}}. [Erratum: {\it Phys.
  Rev. D} {\bf 76}, 129902 (2007)].

\bibitem{Bettoni:2013diz}
D.~Bettoni and S.~Liberati, ``{Disformal invariance of second order
  scalar-tensor theories: Framing the Horndeski action},'' {\em Phys. Rev. D}
  {\bf 88} (2013) 084020, \href{https://arxiv.org/abs/1306.6724}{{\tt
  1306.6724}}.

\bibitem{Takahashi:2021ttd}
K.~Takahashi, H.~Motohashi, and M.~Minamitsuji, ``{Invertible disformal
  transformations with higher derivatives},'' {\em Phys. Rev. D} {\bf 105}
  (2022), no.~2 024015, \href{https://arxiv.org/abs/2111.11634}{{\tt
  2111.11634}}.

\bibitem{Zumalacarregui:2013pma}
M.~Zumalac\'arregui and J.~Garc\'\i{}a-Bellido, ``{Transforming gravity: from
  derivative couplings to matter to second-order scalar-tensor theories beyond
  the Horndeski Lagrangian},'' {\em Phys. Rev. D} {\bf 89} (2014) 064046,
  \href{https://arxiv.org/abs/1308.4685}{{\tt 1308.4685}}.

\bibitem{Domenech:2015tca}
G.~Dom\`enech, S.~Mukohyama, R.~Namba, A.~Naruko, R.~Saitou, and Y.~Watanabe,
  ``{Derivative-dependent metric transformation and physical degrees of
  freedom},'' {\em Phys. Rev. D} {\bf 92} (2015), no.~8 084027,
  \href{https://arxiv.org/abs/1507.05390}{{\tt 1507.05390}}.

\bibitem{Takahashi:2017zgr}
K.~Takahashi, H.~Motohashi, T.~Suyama, and T.~Kobayashi, ``{General invertible
  transformation and physical degrees of freedom},'' {\em Phys. Rev. D} {\bf
  95} (2017), no.~8 084053, \href{https://arxiv.org/abs/1702.01849}{{\tt
  1702.01849}}.

\bibitem{Takahashi:2022mew}
K.~Takahashi, M.~Minamitsuji, and H.~Motohashi, ``{Generalized disformal
  Horndeski theories: cosmological perturbations and consistent matter
  coupling},'' \href{https://arxiv.org/abs/2209.02176}{{\tt 2209.02176}}.

\bibitem{Mukohyama:2022enj}
S.~Mukohyama and V.~Yingcharoenrat, ``{Effective field theory of black hole
  perturbations with timelike scalar profile: formulation},'' {\em JCAP} {\bf
  09} (2022) 010, \href{https://arxiv.org/abs/2204.00228}{{\tt 2204.00228}}.

\bibitem{ArkaniHamed:2003uy}
N.~Arkani-Hamed, H.-C. Cheng, M.~A. Luty, and S.~Mukohyama, ``{Ghost
  condensation and a consistent infrared modification of gravity},'' {\em JHEP}
  {\bf 05} (2004) 074, \href{https://arxiv.org/abs/hep-th/0312099}{{\tt
  hep-th/0312099}}.

\bibitem{Arkani-Hamed:2003juy}
N.~Arkani-Hamed, P.~Creminelli, S.~Mukohyama, and M.~Zaldarriaga, ``{Ghost
  inflation},'' {\em JCAP} {\bf 04} (2004) 001,
  \href{https://arxiv.org/abs/hep-th/0312100}{{\tt hep-th/0312100}}.

\bibitem{Creminelli:2006xe}
P.~Creminelli, M.~A. Luty, A.~Nicolis, and L.~Senatore, ``{Starting the
  Universe: Stable Violation of the Null Energy Condition and Non-standard
  Cosmologies},'' {\em JHEP} {\bf 12} (2006) 080,
  \href{https://arxiv.org/abs/hep-th/0606090}{{\tt hep-th/0606090}}.

\bibitem{Cheung:2007st}
C.~Cheung, P.~Creminelli, A.~L. Fitzpatrick, J.~Kaplan, and L.~Senatore, ``{The
  Effective Field Theory of Inflation},'' {\em JHEP} {\bf 03} (2008) 014,
  \href{https://arxiv.org/abs/0709.0293}{{\tt 0709.0293}}.

\bibitem{Gubitosi:2012hu}
G.~Gubitosi, F.~Piazza, and F.~Vernizzi, ``{The Effective Field Theory of Dark
  Energy},'' {\em JCAP} {\bf 02} (2013) 032,
  \href{https://arxiv.org/abs/1210.0201}{{\tt 1210.0201}}.

\bibitem{Franciolini:2018uyq}
G.~Franciolini, L.~Hui, R.~Penco, L.~Santoni, and E.~Trincherini, ``{Effective
  Field Theory of Black Hole Quasinormal Modes in Scalar-Tensor Theories},''
  {\em JHEP} {\bf 02} (2019) 127, \href{https://arxiv.org/abs/1810.07706}{{\tt
  1810.07706}}.

\bibitem{Hui:2021cpm}
L.~Hui, A.~Podo, L.~Santoni, and E.~Trincherini, ``{Effective Field Theory for
  the perturbations of a slowly rotating black hole},'' {\em JHEP} {\bf 12}
  (2021) 183, \href{https://arxiv.org/abs/2111.02072}{{\tt 2111.02072}}.

\bibitem{Will:2014xja}
C.~M. Will, ``{The Confrontation between General Relativity and Experiment},''
  {\em Living Rev.Rel.} {\bf 17} (2014) 4,
  \href{https://arxiv.org/abs/1403.7377}{{\tt 1403.7377}}.

\bibitem{Ogawa:2015pea}
H.~Ogawa, T.~Kobayashi, and T.~Suyama, ``{Instability of hairy black holes in
  shift-symmetric Horndeski theories},'' {\em Phys. Rev. D} {\bf 93} (2016),
  no.~6 064078, \href{https://arxiv.org/abs/1510.07400}{{\tt 1510.07400}}.

\bibitem{Takahashi:2015pad}
K.~Takahashi, T.~Suyama, and T.~Kobayashi, ``{Universal instability of hairy
  black holes in Lovelock-Galileon theories in D dimensions},'' {\em Phys. Rev.
  D} {\bf 93} (2016), no.~6 064068,
  \href{https://arxiv.org/abs/1511.06083}{{\tt 1511.06083}}.

\bibitem{Takahashi:2016dnv}
K.~Takahashi and T.~Suyama, ``{Linear perturbation analysis of hairy black
  holes in shift-symmetric Horndeski theories: Odd-parity perturbations},''
  {\em Phys. Rev. D} {\bf 95} (2017), no.~2 024034,
  \href{https://arxiv.org/abs/1610.00432}{{\tt 1610.00432}}.

\bibitem{Tretyakova:2017lyg}
D.~A. Tretyakova and K.~Takahashi, ``{Stable black holes in shift-symmetric
  Horndeski theories},'' {\em Class. Quant. Grav.} {\bf 34} (2017), no.~17
  175007, \href{https://arxiv.org/abs/1702.03502}{{\tt 1702.03502}}.

\bibitem{Babichev:2017lmw}
E.~Babichev, C.~Charmousis, G.~Esposito-Far\`ese, and A.~Leh\'ebel,
  ``{Stability of Black Holes and the Speed of Gravitational Waves within
  Self-Tuning Cosmological Models},'' {\em Phys. Rev. Lett.} {\bf 120} (2018),
  no.~24 241101, \href{https://arxiv.org/abs/1712.04398}{{\tt 1712.04398}}.

\bibitem{Babichev:2018uiw}
E.~Babichev, C.~Charmousis, G.~Esposito-Far\`ese, and A.~Leh\'ebel,
  ``{Hamiltonian unboundedness vs stability with an application to Horndeski
  theory},'' {\em Phys. Rev. D} {\bf 98} (2018), no.~10 104050,
  \href{https://arxiv.org/abs/1803.11444}{{\tt 1803.11444}}.

\bibitem{Minamitsuji:2018vuw}
M.~Minamitsuji and H.~Motohashi, ``{Stealth Schwarzschild solution in shift
  symmetry breaking theories},'' {\em Phys. Rev. D} {\bf 98} (2018), no.~8
  084027, \href{https://arxiv.org/abs/1809.06611}{{\tt 1809.06611}}.

\bibitem{Takahashi:2019oxz}
K.~Takahashi, H.~Motohashi, and M.~Minamitsuji, ``{Linear stability analysis of
  hairy black holes in quadratic degenerate higher-order scalar-tensor
  theories: Odd-parity perturbations},'' {\em Phys. Rev. D} {\bf 100} (2019),
  no.~2 024041, \href{https://arxiv.org/abs/1904.03554}{{\tt 1904.03554}}.

\bibitem{deRham:2019gha}
C.~de~Rham and J.~Zhang, ``{Perturbations of stealth black holes in degenerate
  higher-order scalar-tensor theories},'' {\em Phys. Rev. D} {\bf 100} (2019),
  no.~12 124023, \href{https://arxiv.org/abs/1907.00699}{{\tt 1907.00699}}.

\bibitem{Charmousis:2019fre}
C.~Charmousis, M.~Crisostomi, D.~Langlois, and K.~Noui, ``{Perturbations of a
  rotating black hole in DHOST theories},'' {\em Class. Quant. Grav.} {\bf 36}
  (2019), no.~23 235008, \href{https://arxiv.org/abs/1907.02924}{{\tt
  1907.02924}}.

\bibitem{Khoury:2020aya}
J.~Khoury, M.~Trodden, and S.~S.~C. Wong, ``{Existence and instability of hairy
  black holes in shift-symmetric Horndeski theories},'' {\em JCAP} {\bf 11}
  (2020) 044, \href{https://arxiv.org/abs/2007.01320}{{\tt 2007.01320}}.

\bibitem{Tomikawa:2021pca}
K.~Tomikawa and T.~Kobayashi, ``{Perturbations and quasinormal modes of black
  holes with time-dependent scalar hair in shift-symmetric scalar-tensor
  theories},'' {\em Phys. Rev. D} {\bf 103} (2021), no.~8 084041,
  \href{https://arxiv.org/abs/2101.03790}{{\tt 2101.03790}}.

\bibitem{Langlois:2021aji}
D.~Langlois, K.~Noui, and H.~Roussille, ``{Black hole perturbations in modified
  gravity},'' {\em Phys. Rev. D} {\bf 104} (2021), no.~12 124044,
  \href{https://arxiv.org/abs/2103.14750}{{\tt 2103.14750}}.

\bibitem{Langlois:2021xzq}
D.~Langlois, K.~Noui, and H.~Roussille, ``{Asymptotics of linear differential
  systems and application to quasinormal modes of nonrotating black holes},''
  {\em Phys. Rev. D} {\bf 104} (2021), no.~12 124043,
  \href{https://arxiv.org/abs/2103.14744}{{\tt 2103.14744}}.

\bibitem{Takahashi:2021bml}
K.~Takahashi and H.~Motohashi, ``{Black hole perturbations in DHOST theories:
  master variables, gradient instability, and strong coupling},'' {\em JCAP}
  {\bf 08} (2021) 013, \href{https://arxiv.org/abs/2106.07128}{{\tt
  2106.07128}}.

\bibitem{Nakashi:2022wdg}
K.~Nakashi, M.~Kimura, H.~Motohashi, and K.~Takahashi, ``{Black hole
  perturbations in higher-order scalar\textendash{}tensor theories: initial
  value problem and dynamical stability},'' {\em Class. Quant. Grav.} {\bf 39}
  (2022), no.~17 175003, \href{https://arxiv.org/abs/2204.05054}{{\tt
  2204.05054}}.

\bibitem{Langlois:2022ulw}
D.~Langlois, K.~Noui, and H.~Roussille, ``{On the effective metric of axial
  black hole perturbations in DHOST gravity},'' {\em JCAP} {\bf 08} (2022),
  no.~08 040, \href{https://arxiv.org/abs/2205.07746}{{\tt 2205.07746}}.

\bibitem{Motohashi:2019ymr}
H.~Motohashi and S.~Mukohyama, ``{Weakly-coupled stealth solution in scordatura
  degenerate theory},'' {\em JCAP} {\bf 01} (2020) 030,
  \href{https://arxiv.org/abs/1912.00378}{{\tt 1912.00378}}.

\bibitem{Arkani-Hamed:2003pdi}
N.~Arkani-Hamed, H.-C. Cheng, M.~A. Luty, and S.~Mukohyama, ``{Ghost
  condensation and a consistent infrared modification of gravity},'' {\em JHEP}
  {\bf 05} (2004) 074, \href{https://arxiv.org/abs/hep-th/0312099}{{\tt
  hep-th/0312099}}.

\bibitem{Mukohyama:2005rw}
S.~Mukohyama, ``{Black holes in the ghost condensate},'' {\em Phys. Rev. D}
  {\bf 71} (2005) 104019, \href{https://arxiv.org/abs/hep-th/0502189}{{\tt
  hep-th/0502189}}.

\bibitem{Motohashi:2020wxj}
H.~Motohashi and W.~Hu, ``{Effective field theory of degenerate higher-order
  inflation},'' {\em Phys. Rev. D} {\bf 101} (2020) 083531,
  \href{https://arxiv.org/abs/2002.07967}{{\tt 2002.07967}}.

\bibitem{Lemaitre:1933gd}
G.~Lemaitre, ``{The expanding universe},'' {\em Annales Soc. Sci. Bruxelles A}
  {\bf 53} (1933) 51--85.

\bibitem{Takahashi:2020hso}
K.~Takahashi and H.~Motohashi, ``{General Relativity solutions with stealth
  scalar hair in quadratic higher-order scalar-tensor theories},'' {\em JCAP}
  {\bf 06} (2020) 034, \href{https://arxiv.org/abs/2004.03883}{{\tt
  2004.03883}}.

\bibitem{Motohashi:2011pw}
H.~Motohashi and T.~Suyama, ``{Black hole perturbation in parity violating
  gravitational theories},'' {\em Phys. Rev. D} {\bf 84} (2011) 084041,
  \href{https://arxiv.org/abs/1107.3705}{{\tt 1107.3705}}.

\bibitem{Regge:1957td}
T.~Regge and J.~A. Wheeler, ``{Stability of a Schwarzschild singularity},''
  {\em Phys. Rev.} {\bf 108} (1957) 1063--1069.

\bibitem{Afshordi:2006ad}
N.~Afshordi, D.~J.~H. Chung, and G.~Geshnizjani, ``{Cuscuton: A Causal Field
  Theory with an Infinite Speed of Sound},'' {\em Phys. Rev. D} {\bf 75} (2007)
  083513, \href{https://arxiv.org/abs/hep-th/0609150}{{\tt hep-th/0609150}}.

\bibitem{Iyonaga:2018vnu}
A.~Iyonaga, K.~Takahashi, and T.~Kobayashi, ``{Extended Cuscuton:
  Formulation},'' {\em JCAP} {\bf 12} (2018), no.~12 002,
  \href{https://arxiv.org/abs/1809.10935}{{\tt 1809.10935}}.

\bibitem{Iyonaga:2020bmm}
A.~Iyonaga, K.~Takahashi, and T.~Kobayashi, ``{Extended Cuscuton as Dark
  Energy},'' {\em JCAP} {\bf 07} (2020) 004,
  \href{https://arxiv.org/abs/2003.01934}{{\tt 2003.01934}}.

\bibitem{DeFelice:2011ka}
A.~De~Felice, T.~Suyama, and T.~Tanaka, ``{Stability of Schwarzschild-like
  solutions in f(R,G) gravity models},'' {\em Phys. Rev. D} {\bf 83} (2011)
  104035, \href{https://arxiv.org/abs/1102.1521}{{\tt 1102.1521}}.

\bibitem{Creminelli:2017sry}
P.~Creminelli and F.~Vernizzi, ``{Dark Energy after GW170817 and GRB170817A},''
  {\em Phys. Rev. Lett.} {\bf 119} (2017), no.~25 251302,
  \href{https://arxiv.org/abs/1710.05877}{{\tt 1710.05877}}.

\bibitem{Creminelli:2019kjy}
P.~Creminelli, G.~Tambalo, F.~Vernizzi, and V.~Yingcharoenrat, ``{Dark-Energy
  Instabilities induced by Gravitational Waves},'' {\em JCAP} {\bf 05} (2020)
  002, \href{https://arxiv.org/abs/1910.14035}{{\tt 1910.14035}}.

\bibitem{Motohashi:2016prk}
H.~Motohashi, T.~Suyama, and K.~Takahashi, ``{Fundamental theorem on gauge
  fixing at the action level},'' {\em Phys. Rev. D} {\bf 94} (2016), no.~12
  124021, \href{https://arxiv.org/abs/1608.00071}{{\tt 1608.00071}}.

\bibitem{Barausse:2012ny}
E.~Barausse and T.~P. Sotiriou, ``{A no-go theorem for slowly rotating black
  holes in Ho\v{r}ava-Lifshitz gravity},'' {\em Phys. Rev. Lett.} {\bf 109}
  (2012) 181101, \href{https://arxiv.org/abs/1207.6370}{{\tt 1207.6370}}.
  [Erratum: \href{https://doi.org/10.1103/PhysRevLett.110.039902}{{\it Phys.
  Rev. Lett} {\bf 110}, 039902 (2013)}].

\bibitem{Wang:2012at}
A.~Wang, ``{On 'No-go theorem for slowly rotating black holes in
  Ho\v{r}ava-Lifshitz gravity'},'' \href{https://arxiv.org/abs/1212.1040}{{\tt
  1212.1040}}.

\bibitem{Barausse:2012qh}
E.~Barausse and T.~P. Sotiriou, ``{Slowly rotating black holes in
  Ho\v{r}ava-Lifshitz gravity},'' {\em Phys. Rev. D} {\bf 87} (2013) 087504,
  \href{https://arxiv.org/abs/1212.1334}{{\tt 1212.1334}}.

\bibitem{Wang:2012nv}
A.~Wang, ``{Stationary axisymmetric and slowly rotating spacetimes in
  Ho\v{r}ava-lifshitz gravity},'' {\em Phys. Rev. Lett.} {\bf 110} (2013),
  no.~9 091101, \href{https://arxiv.org/abs/1212.1876}{{\tt 1212.1876}}.

\bibitem{Hui:2020xxx}
L.~Hui, A.~Joyce, R.~Penco, L.~Santoni, and A.~R. Solomon, ``{Static response
  and Love numbers of Schwarzschild black holes},'' {\em JCAP} {\bf 04} (2021)
  052, \href{https://arxiv.org/abs/2010.00593}{{\tt 2010.00593}}.

\bibitem{Charalambous:2021mea}
P.~Charalambous, S.~Dubovsky, and M.~M. Ivanov, ``{On the Vanishing of Love
  Numbers for Kerr Black Holes},'' {\em JHEP} {\bf 05} (2021) 038,
  \href{https://arxiv.org/abs/2102.08917}{{\tt 2102.08917}}.

\bibitem{Bernard:2019yfz}
L.~Bernard, ``{Dipolar tidal effects in scalar-tensor theories},'' {\em Phys.
  Rev. D} {\bf 101} (2020), no.~2 021501,
  \href{https://arxiv.org/abs/1906.10735}{{\tt 1906.10735}}.

\bibitem{Gorji:2020bfl}
M.~A. Gorji, H.~Motohashi, and S.~Mukohyama, ``{Stealth dark energy in
  scordatura DHOST theory},'' {\em JCAP} {\bf 03} (2021) 081,
  \href{https://arxiv.org/abs/2009.11606}{{\tt 2009.11606}}.

\bibitem{Gorji:2021isn}
M.~A. Gorji, H.~Motohashi, and S.~Mukohyama, ``{Inflation with $0\leq c_{\rm
  s}\leq 1$},'' {\em JCAP} {\bf 02} (2022), no.~02 030,
  \href{https://arxiv.org/abs/2110.10731}{{\tt 2110.10731}}.

\bibitem{Aoki:2021wew}
K.~Aoki, M.~A. Gorji, S.~Mukohyama, and K.~Takahashi, ``{The effective field
  theory of vector-tensor theories},'' {\em JCAP} {\bf 01} (2022), no.~01 059,
  \href{https://arxiv.org/abs/2111.08119}{{\tt 2111.08119}}.

\bibitem{Aoki:2022_prep}
K.~Aoki, M.~A. Gorji, S.~Mukohyama, K.~Takahashi, and V.~Yingcharoenrat,
  ``{Effective Field Theory of Vector-Tensor Theories on Black Hole
  Background},'' {\em {\textit work in progress}}.

\bibitem{Khoury:2022zor}
J.~Khoury, T.~Noumi, M.~Trodden, and S.~S.~C. Wong, ``{Stability of Hairy Black
  Holes in Shift-Symmetric Scalar-Tensor Theories via the Effective Field
  Theory Approach},'' \href{https://arxiv.org/abs/2208.02823}{{\tt
  2208.02823}}.

\bibitem{Motohashi:2018wdq}
H.~Motohashi and M.~Minamitsuji, ``{General Relativity solutions in modified
  gravity},'' {\em Phys. Lett. B} {\bf 781} (2018) 728--734,
  \href{https://arxiv.org/abs/1804.01731}{{\tt 1804.01731}}.

\bibitem{Motohashi:2019sen}
H.~Motohashi and M.~Minamitsuji, ``{Exact black hole solutions in
  shift-symmetric quadratic degenerate higher-order scalar-tensor theories},''
  {\em Phys. Rev. D} {\bf 99} (2019), no.~6 064040,
  \href{https://arxiv.org/abs/1901.04658}{{\tt 1901.04658}}.

\end{thebibliography}\endgroup

\end{document}